\documentclass[12pt]{article}
\usepackage{amscd,amssymb,amsmath,latexsym,enumerate}
\usepackage[mathscr]{euscript}
\usepackage{epsfig}
\usepackage{fancybox}
\usepackage{verbatim}

\usepackage{color}

\title{Lyapunov spectra for all ten symmetry classes \\
of quasi-one-dimensional disordered systems of non-interacting Fermions}

\author{Andreas~W.~W.~Ludwig$^1$, Hermann Schulz-Baldes$^2$, Michael Stolz$^3$
\\
\\
{\small $^1$ Department of Physics, University of California, Santa Barbara, CA, USA}
\\
{\small $^2$ Department Mathematik, Universit\"at Erlangen-N\"urnberg, Germany}
\\
{\small $^3$ Fakult\"at f\"ur Mathematik, Ruhr-Universit\"at Bochum, Germany}
}

\date{ }

\newtheorem{proposi}{Proposition}
\newtheorem{lemma}{Lemma}

\newcommand{\CM}{{\mathbb C}}
\newcommand{\NM}{{\mathbb N}}
\newcommand{\RM}{{\mathbb R}}
\newcommand{\SM}{{\mathbb S}}

\newcommand{\ZM}{{\mathbb Z}}

\newcommand{\HM}{{\mathbb H}}

\newcommand{\Pp}{{\cal P}}

\newcommand{\EE}{{\bf E}}

\newcommand{\Ff}{{\cal F}}
\newcommand{\Gg}{{\cal G}}

\newcommand{\Uu}{{\cal U}}
\newcommand{\Vv}{{\cal V}}

\newcommand{\Oo}{{\cal O}}
\newcommand{\Tr}{\mbox{\rm Tr}}
\newcommand{\Tt}{{\cal T}}

\newcommand{\one}{{\bf 1}}

\def\f{\mathfrak }
\def\frp{{\f p}}
\def\frm{{\f m}}
\def\frn{{\f n}}

\def\calf{{\mathcal F}}
\def\cala{{\mathcal A}}
\def\on{\operatorname}
\def\tr{\on{Tr}}
\def\tra{^{\prime}}
\begin{document}

\maketitle

%%%%%%%%%%%%%%%%%%%%%%%%%%%%%%%%%%%%%%%%%%%%%%%%%%%%
\begin{abstract}
A random phase property is proposed for products of random matrices drawn from 
any one of the classical groups associated  with the  ten Cartan symmetry classes of non-interacting disordered Fermion systems. It allows to calculate the Lyapunov spectrum explicitly in a perturbative regime. These results apply to quasi-one-dimensional random Dirac operators which can be constructed 
as representatives for each of the ten symmetry classes. For those symmetry classes that correspond to two-dimensional topological insulators or superconductors, the random Dirac operators
describing the one-dimensional boundaries have vanishing Lyapunov exponents and almost surely an absolutely continuous spectrum, reflecting the gapless and conducting nature of the boundary degrees of freedom.

\vspace{.1cm}


\end{abstract}
%%%%%%%%%%%%%%%%%%

%\pacs{ 72.15.Rn, % Localization effects (Anderson or weak localization)
%73.20.Fz, %Weak or Anderson localization
%73.23.-b % Electronic transport in mesoscopic systems
%}

The theory of products of random matrices has successfully been applied to the analysis of  Anderson localization effects in quasi-one-dimensional disordered quantum systems. In the mathematical literature \cite{BL}, particular focus has traditionally been put on time-reversal invariant systems without spin, for which the transfer matrices are in the symplectic group. On the other hand, it is well known from work of Altland and Zirnbauer \cite{AZ, HHZ} (see also \cite{SRFL,RSFL})
that all disordered systems of non-interacting Fermions can be fully classified by their behavior under time-reversal, particle-hole  and so-called
chiral (or `sublattice') symmetries (denoted in the sequel by TRS, PHS and SLS).
There are precisely ten ways a system of non-interacting Fermion can respond to these symmetires, and the resulting ten symmetry classes are in one-to-one correspondence with the ten (non-exceptional) symmetric spaces classified by E. Cartan.
In one dimension this leads to transfer matrices in classical groups
other than the symplectic group  (see Section \ref{sec-reviewclassical} below for terminology), and 
in fact  each of the non-exceptional groups of the Cartan classification appears as the transfer matrix group of an adequate quasi-one-dimensional model.  For instance, an SLS implies implies reflection symmetry of the spectrum about a special energy, and this may occur, e.g., as a consequence of a sublattice structure. The best known example for this is probably the Dyson model with off-diagonal disorder, for which the transfer matrices at zero energy have such a chiral symmetry (it is in the chiral orthogonal class). On the other hand, the Bogoliubov-de Gennes (BdG) classes describe quasi-particles in dirty superconductors. The BdG Hamiltonian describing the quantum dynamics of the Fermionic quasiparticles inside the superconductor possesses by construction a built-in PHS, relating particles to holes. Representatives for all symmetry classes can be constructed in terms of one-dimensional random Dirac operators with the corresponding symmetries \cite{BMSA,BFGM}. The interplay between physical symmetries of these Hamiltonians defining the ten Cartan classes and the symmetries of the transfer matrix group is reviewed in Sections~\ref{sec-sym} and \ref{sec-model} below, see also \cite{CAS,SRFL}. Of course, apart from this physically motivated application, the study of random products in each of the classical groups is of intrinsic mathematical interest as well.

\vspace{.2cm}

The most interesting quantities associated to such random products are the Lyapunov exponents $\gamma_p\geq\gamma_{p+1}$ with $p=1,\ldots,N$, where $N$ is the fixed size of the square random matrices. They are the exponential growth rates of the singular values of the random matrix  products, and their formal definition is recalled in Section~\ref{sec-Lyapspec}. The collection of these exponents constitutes the Lyapunov spectrum.  In the applications to quasi-one-dimensional random quantum systems, the smallest non-negative Lyapunov exponent is the inverse localization length and hence has a direct and important physical interpretation. Its positivity is also a crucial prerequisite for mathematical proofs of Anderson localization \cite{BL}. The remaining part of the Lyapunov spectrum is also of physical interest. Indeed, all singular values enter into the expression for the conductance (calculated for a large, but finite number of random matrices, so that strictly speaking not the Lyapunov exponents themselves are relevant). Furthermore, the equidistance of the Lyapunov spectrum enters for example  into the analysis of the shot noise \cite{BB}. For these reasons, and for its own sake, it is interesting to calculate the Lyapunov spectrum of a given model.

\vspace{.2cm}

In the physics literature, the approach most widely used for the study of the Lyapunov spectrum is based on the Dorokhov-Mello-Pereyra-Kumar (DMPK) equations \cite{Dor,MPK}. These equations are universal (in the sense that they do not contain model dependent information other than the above-mentioned symmetry properties), and their derivation relies on the so-called maximal entropy Ansatz. A recently proposed alternative approach \cite{RS} is closer to the theory of products of random matrices and the numerical transfer matrix method. One considers the random action of the transfer matrices on adequate frames (orthogonal coordinate systems). Each frame specifies an increasing sequence of subspaces, so that the action can also be seen as an action on the collection of compact Grassmannians. For each dimension, this leads to a Markov process which is known to have a unique invariant measure $\nu$, called the Furstenberg measure \cite{Fur,BL}. The {\it random phase property} (RPP) supposes that this measure $\nu$ is given by the geometric invariant measure on the Grassmannian.  In \cite{RS} the RPP was only studied for the three standard symmetry classes, and in the present work this notion is extended to treat the other symmetry classes of BdG and chiral type. Moreover, the formulation of the RPP presented here is of more geometric and conceptual nature than in \cite{RS}. In order to keep the presentation and concepts clear, we choose not to include models with hyperbolic (closed) channels here, even though this would be straightforward along the lines of \cite{RS}. Models for which the RPP holds exactly can easily be constructed, see Section~\ref{sec-Lyapspec}. As discussed in \cite{RS}, the RPP is considerably weaker than the maximal entropy Ansatz, and it is amenable to numerical verification in concrete models. 

\vspace{.2cm}

It is well-known \cite{Fur,BL} that the Lyapunov exponents are without approximation given in terms of $\nu$ by the Furstenberg formula
$$
\gamma_p\;=\;\EE_\nu(f_p)\;,
$$
where $f_p$ is a certain explicitly calculable function on the Grassmannian, which depends also on the distribution of the random matrices. Whenever the RPP holds, so that $\nu$ is given by the geometric invariant measure on the Grassmannian, the calculation of the Lyapunov spectrum thus reduces to the evaluation of the corresponding averages. These geometric invariant measures are given in terms of the Haar measure on the maximal compact subgroups of the given classical groups, and all these compact subgroups are in turn given in terms of the unitary, orthogonal, or symplectic group. While in principle it is known how to calculate these averages \cite{Wei,CS,CSt}, it becomes algebraically feasible only in a perturbative regime of small randomness. This means that the matrices entering in the random products are close to unitary with random deviations of small order $\lambda>0$. In this situation the function $f_p$ can be expanded in $\lambda$, and thus only the second and some fourth moments w.r.t. the geometric measures need to be calculated.  The necessary formulas are collected in the Appendix. The outcome are nice and compact formulas for the Lyapunov spectrum in a perturbative regime. They are listed for each symmetry class separately in the subsections of Section~\ref{sec-Lyapspec}. In the case of a centered random perturbation of strength $\lambda$ they are in all cases given by roughly the same formula 
\begin{equation}
\label{eq-mainres}
\gamma_p\;=\;\lambda^2\;C_N\;\frac{N-\eta \,p\,+\,\eta'}{N}\;+\;\Oo(\lambda^3)\;,
\end{equation}
with $\eta\in\{1,2\}$ and $p$-dependent $\eta'\in\{0,\frac{1}{2},1\}$. The prefactor $C_N$ depends on the symmetry class and on the details of the distribution of the random matrices, see the formulas in Section~\ref{sec-Lyapspec}. It is roughly given by $C_N\sim\frac{1}{N}\sum \mbox{\rm Var}(P_{n,m})$ where the sum runs over $n,m=1,\ldots,N$ and $\mbox{\rm Var}(P_{n,m})$ is the variance of the $(n, m)-$entry of the random perturbation of order $\lambda$. These explicit formulas for the Lyapunov exponents for all ten Cartan symmetry  classes are a first set of main results of the paper. They allow to read off the smallest non-negative Lyapunov exponent and show the equidistance of the Lyapunov spectrum, provided the RPP holds. As already pointed out above, it is possible to construct models for which the RPP holds exactly. This is the case for the quasi-one-dimensional random Dirac operators constructed for every symmetry class  in Section~\ref{sec-model}. 

\vspace{.2cm}

The perturbative formula for $f_p$ leads to numerous terms which have to be combined to the compact formula \eqref{eq-mainres}. Thus its proof is a rather lengthy and tedious algebraic calculation in each symmetry class, carried out in Section~\ref{sec-Lyapspec} based on the formulas in the Appendix. We feel this effort is justified by the results above as well as for the following reasons. In many applications, one is confronted with a weak coupling regime of the disorder. In particular, in systems with only few channels (small $N$), the explicit form of the prefactors is of interest. Often numerical calculations are carried out and the analytical formulas above allow to gauge
the accuracy of the numerical data (remembering that only the elliptic channels are taken into account). If there are discrepancies this is a clear sign that one is confronted with a situation in which the RPP does not hold. Furthermore, the treatment is exhaustive in the sense that it covers all ten symmetry classes. The detailed list of the groups may serve as a reference, as well as the discussions of peculiarities of the symmetry classes (such as symmetry enforced or accidental degeneracies of the Lyapunov spectrum). This also highlights some differences between the symmetry classes. 

\vspace{.2cm}

The second set of main results is connected 
with the possibility that the products of random matrices may have vanishing Lyapunov exponents which are symmetry enforced, namely they result from the defining relations of the classical group considered. The most basic example of this type are products in the group $\mbox{\rm U}(N,M)$ with $M>N$, which have at least $M-N$ vanishing Lyapunov exponents (symmetry class A). This appears in an effective description of the edge modes  of 2D integer quantum Hall insulators in terms of
quasi-one-dimensional random Dirac operators. Subclasses of transfer matrices correspond to a description of edge modes of the 2D  thermal- (symmetry class D) \cite{RG} and of the 2D spin quantum Hall insulator (symmetry class C) \cite{SMF}.  By quasi-one-dimensional Kotani theory \cite{KS} for random Dirac operators \cite{Sun,SS} the vanishing of the $M-N$ Lyapunov exponents implies the existence of absolutely continuous spectrum of multiplicity $M-N$. Under supplementary conditions, the arguments of \cite{SS} allow to show that  there is no singular spectrum. For more realistic models for integer quantum Hall
insulators given by half-space magnetic operators, the stability of the absolutely continuous spectrum under (disordered) perturbations can be demonstrated by using positive commutators resulting from the edge channels \cite{BP,FGW}. 

\vspace{.2cm}

Other interesting cases with vanishing Lyapunov exponents concern quasi-one-dimensional Hamiltonians in the Cartan classes AII and DIII which describe boundary states of two-dimensional quantum spin Hall systems \cite{KM} and time-reversal invariant dirty  superconductors \cite{SRFL}respectively. Both systems possess time-reversal invariance and have spin-orbit interactions. The parity of their channel number $N$ can be understood as a $\ZM_2$-invariant and, if it is non-trivial (namely, $N$ is odd), then there are again at least two vanishing Lyapunov exponents (arising from Kramers' degeneracy). This again leads to delocalization and to absolutely continuous spectral measures. The case AII was studied in detail in \cite{SS}. Here in Section~\ref{sec-deloc} we discuss how these results translate to the case DIII, and provide details on the effective models for quantum Hall systems described above. Since the Cartan symmetry classes A, D, C, AII, and DIII correspond precisely\cite{SRFL} to the five topological insulators (superconductors) in two spatial dimensions, and have therefore gapless (`conducting') boundary modes which are protected from Anderson localization\cite{SRFL,RSFL}, while on the other hand, the remaining five Cartan symmetry classes can be Anderson localized by disorder, the vanishing of the Lyapunov exponents
in the first five classes reflects precisely the existence of 2D topological insulators (superconductors) in these symmetry classes.

%%%%%%%%%%%%%%%%%%%%%%%%%%%%%%%%%%%%%%%%%%%%%%%%
\section{A list of the classical groups}
\label{sec-reviewclassical}

The classical groups in the sense of Weyl incorporate certain families of matrix groups. The list is given in the first chapter of \cite{GW} and recapitulated below. There is also a close connection with the Cartan classification of symmetric spaces, because each non-compact classical group from the list below corresponds to a non-compact Lie algebra in duality with a classical compact symmetric space \cite{Hel}. As such, there is a Cartan label associated to each classical group, and this label has been used heavily in the physics community \cite{AZ,BFGM,SRFL}. For this reason, we also choose to attach the Cartan label to each classical group $\Gg$, but at the same time also the classical terminology is recalled in each case below. Because it will be heavily used later on, we also recall the maximal compact subgroup $\Uu$ for each classical group $\Gg$, as well as peculiarities of the spectral theory of elements in $\Gg$. The groups $\Gg$ with their Cartan label are listed in Table~1, albeit in a different order than we go through them here. The groups will be described using the matrices
\begin{equation}
\label{eq-symmdef}
J
\;=\;
\begin{pmatrix}
\one & 0 \\ 0 & -\one
\end{pmatrix}
\;,
\qquad
I
\;=\;
\begin{pmatrix}
0 & -\one \\ \one & 0
\end{pmatrix}
\;,
\qquad
K
\;=\;
\begin{pmatrix}
0 & \one \\ \one & 0
\end{pmatrix}
\;,
\end{equation}
where the identity $\one$ is of adequate size, mostly an $N\times N$ matrix. These are real matrices satisfying $J^*=J$, $K^*=K$, $I^*=-I$ and $\one=J^2=-I^2=K^2$, which is why we call $J$ and $K$ an even, but $I$ an odd symmetry. In these formulas and below, $A^*=\overline{A}^t$ denotes the adjoint given by the conjugate-transpose. The matrices $J$, $I$ and $K$ do not commute, and one has $KJ=I$. Of course, one could use Pauli matrices instead, but they are not real, and this is notationally less convenient  for the discussion below. Writing $\imath = \sqrt{-1}$, the three involutions $J$, $K$ and $\imath\,I$ are isospectral, and one can pass from one to another by a unitary transformation. In particular, the Cayley transformation $C$ achieves the following:
\begin{equation}
\label{eq-Cayley}
C^*\,J\,C\;=\;\imath\,I\,
\;,
\qquad
C\,J\,C^*\;=\;K\,
\;,\qquad
{\rm where}\quad
C
\;=\;
\frac{1}{\sqrt{2}}\,
\begin{pmatrix}
\one & -\imath\,\one \\ \one & \imath\,\one
\end{pmatrix}
\;.
\end{equation}
Moreover, $C^*IC=\imath K$, $C^tKC=\one$, $CKC^t=\imath J$ and $CIC^t=-\imath I$. Among the possible realizations of the classical groups, we choose ones for which the maximal compact subgroup takes a particularly simple form, since this will allow us to simplify the calculation of averages over these subgroups later on.

%%%%%%%%%%%%%%%%%%%%%%%%%%%%%%%%%%%%%%%%%%%%%
\subsection{Class A}
\label{sec-GA}

In Class A there are no further symmetries imposed. Thus $\Gg^{\mbox{\rm\tiny A}}=\mbox{\rm GL}(N,\CM)$ is the set of all invertible $N\times N$ matrices. The maximal compact subgroup $\Uu^{\mbox{\rm\tiny A}}=\mbox{\rm U}(N)$ is given by the unitary matrices.

%%%%%%%%%%%%%%%%%%%%%%%%%%%%%%%%%%%%%%%%%%%%%
\subsection{Class AI}
\label{sec-GAI}

Class AI contains the subgroup of $\Gg^{\mbox{\rm\tiny A}}$ composed of the real matrices (invariant under the operation of complex conjugation):
$$
\Gg^{\mbox{\rm\tiny AI}}
\;=\;
\left\{
T\in\mbox{\rm GL}(N,\CM)
\,\left|\,\overline{T}=T\,\right.\right\}
\;=\;
\mbox{\rm GL}(N,\RM)
\;.
$$
Clearly the maximal compact subgroup $\Uu^{\mbox{\rm\tiny AI}}=\mbox{\rm O}(N)$ is composed of the orthogonal matrices. 

%%%%%%%%%%%%%%%%%%%%%%%%%%%%%%%%%%%%%%%%%%%%%
\subsection{Class AII}
\label{sec-GAII}

The definition of Class AII is
$$
\Gg^{\mbox{\rm\tiny AII}}
\;=\;
\left\{
T\in\mbox{\rm GL}(2N,\CM)
\,\left|\,I^*\overline{T}I=T\,\right.\right\}
\;=\;
\mbox{\rm U}^*(2N)
\;.
$$
Writing out the symmetry one readily checks that $T\in\mbox{\rm U}^*(2N)$ is of the form $T=\binom{A\;\;B}{-\overline{B}\;\overline{A}}$ with entries given by $N\times N$ complex matrices. One obtains an $N\times N$ quaternion matrix from $T$ by setting $\Ff(T)=A+Bj$ where $j$ is the  second of the standard unit quaternions $\imath, j,k$ satisfying $\imath^2=j^2=k^2=\imath jk=-1$. This map $\Ff$ is a $*$-isomorphism from $\mbox{\rm U}^*(2N)$ to the group $\mbox{\rm GL}(N,\HM)$ of invertible quaternion matrices. Thus the classes A, AI, AII simply correspond to the invertible matrices over the fields $\CM,\RM,\HM$, respectively. The maximal compact subgroup of $\mbox{\rm U}^*(2N)$ is the symplectic group
$$
\Uu^{\mbox{\rm\tiny AII}}
\;=\;
\mbox{\rm U}^*(2N)
\,\cap\,
\mbox{\rm U}(2N)
\;=\;
\left\{
\left.
\begin{pmatrix}
A & B \\ -\overline{B} & \overline{A}
\end{pmatrix}
\,\right|\,
A^*A+B^t\overline{B}=\one\,,\;\;A^*B^t=B^t\overline{A}\,
\right\}
\;=\;
\mbox{\rm SP}(2N)
\;.
$$
The spectral theory of $T\in\mbox{\rm U}^*(2N)$ has the following feature. If $Tv=\lambda v$ for some $v\in\CM^{2N}$ and $\lambda\in\CM$, then $T\,I\overline{v}=\overline{\lambda}\,I\overline{v}$, and furthermore the vectors $v$ and $I\overline{v}$ are linearly independent. Indeed, suppose $v=\binom{a}{b}$ satisfies $v=\mu I\overline{v}$ for some non-vanishing $\mu\in\CM$. Then $a=-\mu\overline{b}$ and $b=\mu\overline{a}$, which combined imply $a=-|\mu|^2a$ so that $a=b=0$. This implies that the real spectrum of $T\in\mbox{\rm U}^*(2N)$ has even geometric multiplicity, a fact that  is called Kramers' degeneracy in the physics literature. Actually, all the above also holds for all (possibly non-invertible) matrices in the Lie algebra $\mbox{\rm u}^*(2N)$ (which by the above map $\Ff$ is mapped to all $N\times N$ quaternion matrices). 

%%%%%%%%%%%%%%%%%%%%%%%%%%%%%%%%%%%%%%%%%%%%%
\subsection{Class AIII}
\label{sec-GAIII}

The group of Class AIII is defined by
$$
\Gg^{\mbox{\rm\tiny AIII}}
\;=\;
\left\{
T\in\mbox{\rm GL}(2N,\CM)
\,\left|\,T^*\,J\,T=J\,\right.\right\}
\;=\;
\mbox{\rm U}(N,N)
\;,
$$
for which the maximal compact subgroup is
\begin{comment}
the Cayley transform is
%
\begin{equation}
\label{eq-CaleyUNN}
C^*\,\Gg^{\mbox{\rm\tiny AIII}}\,C
\;=\;
\left\{
T\in\mbox{\rm GL}(2N,\CM)
\,\left|\,T^*\,I\,T=I\,\right.\right\}
\;.
\end{equation}
%
\end{comment}
%
\begin{equation}
\label{M-AIIIlabel}
\Uu^{\mbox{\rm\tiny AIII}}
\;=\;
\mbox{\rm U}(N,N)
\,\cap\,
\mbox{\rm U}(2N)
\;=\;
\left\{
\left.
\begin{pmatrix}
V & 0 \\ 0 & W
\end{pmatrix}
\,\right|\,
V,W\in\mbox{\rm U}(N)
\right\}
\;=\;
\mbox{\rm U}(N)\times \mbox{\rm U}(N)
\;.
\end{equation}
Elements of $\mbox{\rm U}(N,N)$ are often also called $J$-unitaries because they conserve the quadratic form $J$. Their spectrum has a reflection property around the unit circle $\SM^1$ because $T-\lambda\one= -\lambda J(T^*)^{-1}(T-(\overline{\lambda})^{-1}\one)^*J$ 
so that the invertibility of $T-\lambda\one$ is indeed equivalent to the invertibility of $T-(\overline{\lambda})^{-1}\one$. Note, however, that in general the two eigenvectors satisfy no relation so that one also cannot conclude that each eigenvalue on the unit circle is degenerate. Another important feature of the spectral theory in $\mbox{\rm U}(N,N)$ is that the eigenvectors $v$ and $w$ corresponding to eigenvalues $\lambda$ and $\mu$ with $\overline{\mu}\lambda\not = 1$ are $J$-orthogonal, namely $w^*Jv=0$. Indeed, $Tv=\lambda v$ and $Tw=\mu w$ imply $w^*Jv=(\mu^{-1}Tw)^*J\lambda^{-1}Tv= (\overline{\mu}\lambda)^{-1} w^*Jv$ which implies the claim.  For the generalized eigenvectors the same $J$-orthogonality holds. Let us sketch the argument for the generalized eigenvector $v'$ of first degree, namely satisfying $Tv'=\lambda v'+v$. As $w^*Jv=0$, one now has $w^*Jv'=(\lambda)^{-1}w^*J(Tv'-v)=(\lambda)^{-1}w^*JTv'=(\overline{\mu}\lambda)^{-1}w^*Jv'=0$. Iteration allows to analyze generalized eigenvectors of higher degree in a similar manner, as well as the $J$-orthogonality between all generalized eigenvectors corresponding to eigenvalues $\lambda$ and $\mu$ satisfying $\overline{\mu}\lambda\not=1$. An alternative argument can be given using Riesz projections. Note that, in particular, this shows that eigenvectors $v$ corresponding to eigenvalues off the unit circle are $J$-isotropic, that is, they satisfy $v^*Jv=0$. 

\vspace{.2cm}

Class AIII also contains a generalization of the group $\Gg^{\mbox{\rm\tiny AIII}}$ defined above:
\begin{equation}
\label{eq-UNM}
\mbox{\rm U}(N,M)
\;=\;
\left\{
T\in\mbox{\rm GL}(N+M,\CM)
\,\left|\,T^*\,G\,T=G\,\right.\right\}
\;,
\qquad
{\rm where}\quad
G
\;=\;
\begin{pmatrix}
\one_N & 0 \\ 0 & -\,\one_M
\end{pmatrix}
\;,
\end{equation}
and $M\geq N$ (say) are possibly different. Then  $\mbox{\rm U}(N,M)\,\cap\,\mbox{\rm U}(N+M)=\mbox{\rm U}(N)\times \mbox{\rm U}(M)$ is the maximal compact subgroup. All the above arguments about the spectral theory still apply to $T\in\mbox{\rm U}(N,M)$. In particular, the spectrum always comes in pairs $\lambda,(\overline{\lambda})^{-1}$, and generalized eigenvectors corresponding to eigenvalues $\lambda$ and $\mu$ with $\overline{\mu}\lambda\not = 1$ are $J$-orthogonal. This implies that there are at most $N$ eigenvalues off the unit circle, or, alternatively stated, at least $M-N$ eigenvalues on the unit circle (here both $N$ and $M-N$ give the algebraic multiplicity). Indeed, let $v_1,\ldots,v_p$ be a maximal set of linear independent generalized eigenvectors of eigenvalues of modulus larger than $1$. By the above,  they then span a $p$-dimensional $G$-isotropic subspace, namely a subspace on which $G$ seen as quadratic form vanishes. But the maximal dimension of $G$-isotropic subspaces is $N$, so that $p\leq N$.

%%%%%%%%%%%%%%%%%%%%%%%%%%%%%%%%%%%%%%%%%%%%%
\subsection{Class CI}
\label{sec-GCI}

Class CI consists of the following subgroup of $\Gg^{\mbox{\rm\tiny AIII}}=\mbox{\rm U}(N,N)$:
$$
\Gg^{\mbox{\rm\tiny CI}}
\;=\;
\left\{
T\in\mbox{\rm GL}(2N,\CM)
\,\left|\,T^*\,J\,T=J\,,\;\;K^*\,\overline{T}\,K=T\right.\right\}
\;.
$$
Its maximal compact subgroup is
\begin{equation}
\label{eq-UCI}
\Uu^{\mbox{\rm\tiny CI}}
\;=\;
\Gg^{\mbox{\rm\tiny CI}}
\,\cap\,
\mbox{\rm U}(2N)
\;=\;
\left\{
\left.
\begin{pmatrix}
V & 0 \\ 0 & \overline{V}
\end{pmatrix}
\,\right|\,
V\in\mbox{\rm U}(N)
\right\}
\;\cong\;
\mbox{\rm U}(N)
\;.
\end{equation}
The group $\Gg^{\mbox{\rm\tiny CI}}$ is actually a disguised version of the real non-compact symplectic group, namely using the Cayley transform one has
$$
C^*\,\Gg^{\mbox{\rm\tiny CI}}\,C
\;=\;
\left\{
T\in\mbox{\rm GL}(2N,\CM)
\,\left|\,T^*\,I\,T=I\,,\;\;\overline{T}=T\right.\right\}
\;=\;
\mbox{\rm SP}(2N,\RM)
\;.
$$
In this representation the maximal compact subgroup does not take a form as simple as \eqref{eq-UCI}. The spectrum of $T\in\Gg^{\mbox{\rm\tiny CI}}$ has, of course, the $\SM^1$-reflection property holding for any matrix in $\Gg^{\mbox{\rm\tiny AIII}}$, but, moreover, $Tv=\lambda v$ leads to $T\,K\overline{v}=\overline{\lambda}\,K\overline{v}$. Thus the eigenvalues come in quadruples $\lambda,\overline{\lambda},\lambda^{-1},(\overline{\lambda})^{-1}$. Let us point out that $v$ and $K\overline{v}$  are, in general, not linearly independent. Therefore one cannot conclude from the above that real eigenvalues are degenerate.

%%%%%%%%%%%%%%%%%%%%%%%%%%%%%%%%%%%%%%%%%%%%%
\subsection{Class DIII}
\label{sec-GDIII}

Class DIII is also a subgroup of $\Gg^{\mbox{\rm\tiny AIII}}=\mbox{\rm U}(N,N)$:
$$
\Gg^{\mbox{\rm\tiny DIII}}
\;=\;
\left\{
T\in\mbox{\rm GL}(2N,\CM)
\,\left|\,T^*\,J\,T=J\,,\;\;I^*\,\overline{T}\,I=T\right.\right\}
\;.
$$
Thus $\Gg^{\mbox{\rm\tiny DIII}}=\mbox{\rm U}(N,N)\,\cap\,\mbox{\rm U}^*(2N)$, and one can also view $\Gg^{\mbox{\rm\tiny DIII}}$ as a subgroup of $\Gg^{\mbox{\rm\tiny AII}}$. In particular, the map $\Ff$ described in Section~\ref{sec-GAII} is well-defined on $\Gg^{\mbox{\rm\tiny DIII}}$, and the image $\Ff(\Gg^{\mbox{\rm\tiny DIII}})$ is given by those quaternion matrices $\Ff(T)$ satisfying $\Ff(T)^*\imath\,\Ff(T)=\imath\,\one$. Its maximal compact subgroup is
$$
\Uu^{\mbox{\rm\tiny DIII}}
\;=\;
\Gg^{\mbox{\rm\tiny DIII}}
\,\cap\,
\mbox{\rm U}(2N)
\;=\;
\left\{
\left.
\begin{pmatrix}
V & 0 \\ 0 & \overline{V}
\end{pmatrix}
\,\right|\,
V\in\mbox{\rm U}(N)
\right\}
\;\cong\;
\mbox{\rm U}(N)
\;.
$$
Taking the Cayley transform of $\Gg^{\mbox{\rm\tiny DIII}}$ one finds, using $C^tIC=\imath\,I$,  another one of the classical groups:
$$
C^*\,\Gg^{\mbox{\rm\tiny DIII}}\,C
\;=\;
\left\{
T\in\mbox{\rm GL}(2N,\CM)
\,\left|\,T^*\,I\,T=I\,,\;\;
I^*\overline{T}I=T
\,\right.\right\}
\;=\;
\mbox{\rm SO}^*(2N)
\;.
$$
Here the equation $I^*\overline{T}I=T$  can equivalently be replaced by $T^tT=\one$, because the relation $T^*IT=I$ holds. As to the spectral theory, again the $\SM^1$-reflection property holds. Furthermore, $Tv=\lambda v$ implies $T\,I\overline{v}=\overline{\lambda}\,I\overline{v}$ so that again the spectrum comes in quadruples as in $\Gg^{\mbox{\rm\tiny CI}}\cong\mbox{\rm SP}(2N,\RM)$. Furthermore, because $\Gg^{\mbox{\rm\tiny DIII}}\subset \Gg^{\mbox{\rm\tiny AII}}$, the real spectrum has again a Kramers' degeneracy (see Section~\ref{sec-GAII}).

\vspace{.2cm}

It is worth mentioning that the two symmetries $J$ and $I$ used in the definition of $\Gg^{\mbox{\rm\tiny DIII}}$ do not commute. If the matrix size is $4N$ instead of $2N$,  which we realize by tensorizing with $\CM^2$ (or equivalently, $N$ is even in the above), then one can also use two commuting symmetries:
$$
(CB)^*\Gg^{\mbox{\rm\tiny DIII}} CB
\;=\;
\left\{
T\in\mbox{\rm GL}(4N,\CM)
\,\left|\,T^*\,I\otimes\one \,T=I\otimes\one\,,\;\;\one\otimes I^*\,\overline{T}\,\one\otimes I=T\right.\right\}
\;,
$$
where $B=2^{-\frac{1}{2}}\binom{-\imath I\;\;\one}{\,\one\;\;\imath I}$. Indeed, then $B^*=B^{-1}=B$ as well as  $\overline{B}\,\one\otimes I B=-\imath \,I\otimes \one$ and $B^* I\otimes\one B=-I\otimes \one$. This representation of the group $\Gg^{\mbox{\rm\tiny DIII}}$ appears as transfer matrices in physical models with odd spin, where then $\one\otimes I$ acts on the spin degree of freedom only (e.g. \cite{RS}). 

%%%%%%%%%%%%%%%%%%%%%%%%%%%%%%%%%%%%%%%%%%%%%
\subsection{Class BDI}
\label{sec-GBDI}

Class BDI is yet another subgroup of $\Gg^{\mbox{\rm\tiny AIII}}=\mbox{\rm U}(N,N)$:
$$
\Gg^{\mbox{\rm\tiny BDI}}
\;=\;
\left\{
T\in\mbox{\rm GL}(2N,\CM)
\,\left|\,T^*\,J\,T=J\,,\;\;\overline{T}=T\,\right.\right\}
\;=\;
\mbox{\rm O}(N,N)\;.
$$
Of course, $\Gg^{\mbox{\rm\tiny BDI}}$ is also the subgroup of $\Gg^{\mbox{\rm\tiny AI}}=\mbox{\rm GL}(2N,\RM)$ specified by the relation $T^tJT=J$. Its maximal compact subgroup is
$$
\Uu^{\mbox{\rm\tiny BDI}}
\;=\;
\Gg^{\mbox{\rm\tiny BDI}}
\,\cap\,
\mbox{\rm U}(2N)
\;=\;
\left\{
\left.
\begin{pmatrix}
V & 0 \\ 0 & W
\end{pmatrix}
\,\right|\,
V,W\in\mbox{\rm O}(N)
\right\}
\;=\;
\mbox{\rm O}(N)
\times \mbox{\rm O}(N)
\;.
$$
As matrices in $\mbox{\rm O}(N,N)$ are real, $\lambda$ is an eigenvalue if and only if $\overline{\lambda}$ is. Therefore, the spectrum for matrices in $\mbox{\rm O}(N,N)$ comes again as complex quadruples, or real pairs $\lambda,\lambda^{-1}$. These pairs are in general not degenerate though (an example for a matrix in $\mbox{\rm O}(1,1)$ with simple real eigenvalue pairs is the hyperbolic rotation). Similarly as in \eqref{eq-UNM}, it is possible to define the groups $\mbox{\rm O}(N,M)$ with maximal compact subgroup $\mbox{\rm O}(N)\times \mbox{\rm O}(M)$. Similar spectral properties as for $\mbox{\rm U}(N,M)$ hold, but this will not be spelled out in detail.

%%%%%%%%%%%%%%%%%%%%%%%%%%%%%%%%%%%%%%%%%%%%%
\subsection{Class CII}
\label{sec-GCII}

Class CII is also a subgroup of $\Gg^{\mbox{\rm\tiny AIII}}$, however, with a symmetry that requires tensorizing with another $\CM^2$ so that the matrices are of size $4N\times 4N$. The symmetry in $\mbox{\rm U}(2N,2N)$ is denoted by $J\otimes\one$ instead of just $J$ in order to highlight the difference. Then
$$
\Gg^{\mbox{\rm\tiny CII}}
\;=\;
\left\{
T\in\mbox{\rm GL}(4N,\CM)
\,\left|\,T^*\,J\otimes\one\,T=J\otimes\one \,,\;\;J\otimes I^*\,\overline{T}\,J\otimes I=T\,\right.\right\}
\,=\,
\mbox{\rm SP}(2N,2N)
\;,
$$
and
$$
\Uu^{\mbox{\rm\tiny CII}}
\;=\;
\Gg^{\mbox{\rm\tiny CII}}
\,\cap\,
\mbox{\rm U}(4N)
\;=\;
\left\{
\left.
\begin{pmatrix}
V & 0 \\ 0 & W
\end{pmatrix}
\,\right|\,
V,W\in\mbox{\rm SP}(2N)
\right\}
\;=\;
\mbox{\rm SP}(2N)
\times \mbox{\rm SP}(2N)
\;.
$$
Obviously, $\Gg^{\mbox{\rm\tiny CII}}\subset \mbox{\rm U}(2N,2N)$, but one also has
$$
D^*\,\Gg^{\mbox{\rm\tiny CII}}\,D
\;=\;
\left\{
T\in\mbox{\rm GL}(4N,\CM)
\,\left|\,T^*\,I\otimes I\,T=I\otimes I \,,\;\;I^*\otimes \one\,\overline{T}\,I\otimes \one=T\,\right.\right\},
$$ 
where $D=2^{-\frac{1}{2}}\binom{\one \; I}{I\;\one}$, and this shows that $D^*\Gg^{\mbox{\rm\tiny CII}}D\subset \mbox{\rm U}^*(4N)$. It is not possible, however,  to write $\Gg^{\mbox{\rm\tiny CII}}$ as an intersection of $\mbox{\rm U}(2N,2N)$ and $\mbox{\rm U}^*(4N)$ (which was possible in Class DIII).

\vspace{.2cm}

By the same reasoning as in Class DIII (see Section~\ref{sec-GDIII}), the spectrum of $T\in\mbox{\rm SP}(2N,2N)$ comes in quadruples, and real eigenvalues always have even geometric multiplicity. However, $Tv=\lambda v$ here implies $T\,J\otimes I\, \overline{v}=\overline{\lambda}\,J\otimes I \,\overline{v}$. Class DIII has again a generalization $\mbox{\rm SP}(2N,2M)$ as in \eqref{eq-UNM}, but the obvious details are not given here.

%%%%%%%%%%%%%%%%%%%%%%%%%%%%%%%%%%%%%%%%%%%%%
\subsection{Class D}
\label{sec-GD}

Class D is again a subgroup of $\Gg^{\mbox{\rm\tiny A}}$ of matrices of size $N\times N$:
$$
\Gg^{\mbox{\rm\tiny D}}
\;=\;
\left\{
T\in\mbox{\rm GL}(N,\CM)
\,\left|\,T^tT=\one\,\right.\right\}
\;=\;
\mbox{\rm O}(N,\CM)
\;.
$$
Note that this is also a $*$-group, namely $T^*\in\mbox{\rm O}(N,\CM)$ if and only if $T\in\mbox{\rm O}(N,\CM)$. Clearly the maximal compact subgroup is $\Uu^{\mbox{\rm\tiny D}}=\Gg^{\mbox{\rm\tiny D}}\cap \mbox{\rm U}(N)= \mbox{\rm O}(N)$. The relation $T^tT=\one$ implies that the spectrum of $T^*$ is given by the $\SM^1$-reflection $\lambda\mapsto (\overline{\lambda})^{-1}$of the spectrum of $T$. Therefore the spectrum of a self-adjoint element $T=T^*\in \mbox{\rm O}(N,\CM)$ is invariant under the map $\lambda\mapsto \lambda^{-1}$.

%%%%%%%%%%%%%%%%%%%%%%%%%%%%%%%%%%%%%%%%%%%%%
\subsection{Class C}
\label{sec-GC}

Finally Class C is also a subgroup of $\Gg^{\mbox{\rm\tiny A}}$, albeit of size $2N\times 2N$:
$$
\Gg^{\mbox{\rm\tiny C}}
\;=\;
\left\{
T\in\mbox{\rm GL}(2N,\CM)
\,\left|\,T^tIT=I\,\right.\right\}
\;=\;
\mbox{\rm SP}(2N,\CM)
\;.
$$
This is also a $*$-group. The maximal compact subgroup is $\Uu^{\mbox{\rm\tiny C}}=\Gg^{\mbox{\rm\tiny C}}\cap \mbox{\rm U}(2N)= \mbox{\rm SP}(2N)$. The conclusions for the spectral theory of $T\in\mbox{\rm SP}(2N,\CM)$ are the same as in Class D.

%%%%%%%%%%%%%%%%%%%%%%%%%%%%%%%%%%%%%%%%%%%%%%%%
\section{Lyapunov spectrum with random phase property}
\label{sec-Lyapspec}

Consider a sequence $(T_n)_{n\geq 0}$ of independently drawn and identically distributed random matrices in one of the above classical groups $\Gg\subset\mbox{\rm GL}(N,\CM)$. For sake of simplicity, these distributions are supposed to have finite moments (existence of third moments is sufficient for the second order perturbation theory later on). One is then interested in the growth in $n$ of the norm of the products $T_n\cdots T_1$, or what is equivalent, the growth of their largest singular value. As one might expect, it turns out that this growth is exponential with a rate defining the top Lyapunov exponent: 
$$
\gamma_1
\;=\;
\lim_{n\to\infty}\;\frac{1}{n}\;\log
\left(
\|T_n\cdots T_1\|
\right)
\;.
$$

It is known \cite{BL} that the limit is almost surely constant and therefore an expectation can be introduced on the r.h.s. before the limit is taken.  Furthermore, instead of calculating an operator norm, an arbitrary vector $v$, say of unit length, may be inserted so that one actually only needs to calculate the length of vectors: 
$$
\gamma_1
\;=\;
\lim_{n\to\infty}\;\frac{1}{n}\;
\EE\;\log
\left(
\|T_n\cdots T_1 v\|
\right)
\;.
$$
Of course, also the growth rates $\gamma_p$ of the other singular values are of interest.  Together they constitute the Lyapunov spectrum. One convenient way to extract them is to  use exterior powers $\Lambda^pT_N\cdots T_1$ acting on the $p$-dimensional Grassmannian identified with the decomposable unit vectors in $\Lambda^p \CM^N$, where $N$ is the size  of the matrices \cite{BL}. Then
$$
\sum_{q=1}^p\gamma_q
\;=\;
\lim_{n\to\infty}\;\frac{1}{n}\;
\EE\;\log
\left(
\|\Lambda^pT_n\cdots T_1\, v_1\wedge\ldots\wedge v_p\|
\right)
\;,
$$
where $v_1,\ldots,v_p$ are unit vectors in $\CM^N$ which are at least orthogonal (further conditions will be imposed below when $\Gg$ is not $\Gg^{\mbox{\rm\tiny A}}$), and the norm is now calculated in $\Lambda^p\CM^N$. It is now very convenient to telescope the r.h.s. so that it becomes a Birkhoff sum. For that purpose let us introduce a sequence $(v^{(p)}(n))_{n\geq 0}$ of decomposable unit vectors in $\Lambda^p \CM^N$ by
\begin{equation}
\label{eq-randdyn}
v^{(p)}(n)
\;=\;
\frac{\Lambda^p T_nv^{(p)}(n-1)}{\| \Lambda^p T_nv^{(p)}(n-1)\|}\;,
\qquad
v^{(p)}(0)=v_1\wedge\ldots\wedge v_p
\;.
\end{equation}
Indeed, then one has
$$
\sum_{q=1}^p\gamma_q
\;=\;
\lim_{n\to\infty}\;\frac{1}{n}\;
\sum_{k=1}^n\;
\EE\;
\log
\left(
\|\Lambda^pT_k\,v^{(p)}(k-1)\|
\right)
\;.
$$
As the conditional expectation over $T_k$ can immediately be taken in each summand (because $v^{(p)}(k-1)$ is independent of $T_k$), one is therefore confronted with the calculation of Birkhoff sums of the random dynamical system \eqref{eq-randdyn} on the $p$-dimensional Grassmannian in $\CM^N$, which is a compact Riemannian manifold. It is known \cite{BL} that this Markov process has a unique invariant and ergodic measure $\nu$ called the Furstenberg measure. In terms of expectations w.r.t. $\nu$, one then has the so-called Furstenberg formula \cite{Fur}:
\begin{equation}
\label{eq-Furstenberg}
\sum_{q=1}^p\gamma_q
\;=\;
\EE_T\;\EE_\nu\;
\log
\left(
\|\Lambda^pT\,v^{(p)}\|
\right)
\;.
\end{equation}
The measure $\nu$ is always H\"older continuous, and, if the distribution of $T$ is absolutely continuous, then also $\nu$ is absolutely continuous \cite{BL}. Here we will be particularly interested in a situation where $T_n$ is of the form
\begin{equation}
\label{eq-perturbform}
T_n\;=\;R_n\,e^{\lambda P_n}
\;,
\end{equation}
where $R_n\in\Gg$ is unitary, $\lambda$ is a real coupling constant, and $P_n$ is in the Lie algebra of $\Gg$. In applications, the random matrices $T_n$ may be of the form \eqref{eq-perturbform} only after an adequate basis transformation independent of $n$. An example of this type are discrete random Schr\"odinger operators \cite{RS0,RS}. If the distribution of the $R_n$ and $P_n$ satisfies a certain coupling hypothesis formulated in \cite{SS0} (of H\"ormander type), then the invariant measure $\nu=\nu_\lambda$ is in a weak sense and with errors of order $\lambda$ absolutely continuous w.r.t.\ the invariant distribution \cite{SS0}. This invariant distribution is simply obtained by taking a decomposable unit vector $v^{(p)}\in\Lambda^p\CM^N$ and randomly rotating it by $\Lambda^pU$, with the unitary $U$ drawn according to the Haar measure on the maximal compact subgroup $\Uu\subset\Gg$. If $\Gg=\Gg^{\mbox{\rm\tiny A}}=\mbox{\rm GL}(N,\CM)$, the measure on the $p$-Grassmannian (still identified with the decomposable unit vectors in $\Lambda^p\CM^N$) is independent of the choice of the reference vector $v^{(p)}$. However, if $\Gg$ is not equal to $\Gg^{\mbox{\rm\tiny A}}=\mbox{\rm GL}(N,\CM)$ so that $\Uu$ is smaller than $\Uu^{\mbox{\rm\tiny A}}=\mbox{\rm U}(N)$, the distrubution thus obtained need not be supported on the full Grassmannian. Thus in this case the choice of the reference vector $v^{(p)}$ becomes relevant. The guiding principle in determining the correct choice is that the vectors $v_q$ composing $v^{(p)}$ as in \eqref{eq-randdyn} {\it can} be the eigenvectors of the a positive element in $\Gg$ (which is $(T_N\cdots T_1)^*T_N\cdots T_1$ in the application above). For instance, if $\Gg=\mbox{\rm U}(N,N)$, this imposes that the $v_q$ are $J$-isotropic and pairwise $J$-orthogonal. Furthermore, if the spectral theory in $\Gg$ leads to Kramers' degeneracy, then also the vectors $v_q$ have to respect that structure and come in adequate pairs. The choice of the orthonormal basis will be discussed in detail in each case below. Apart from this (admittedly important) detail, we can now state the central hypothesis of this paper extending that of \cite{RS}. Strictly speaking, it extends \cite{RS}  to other symmetry classes, but is less general because no hyperbolic channels in the terminology of \cite{RS0,RS} are considered here.

\vspace{.2cm}

\noindent {\bf Random phase property (RPP):} {\sl The invariant measure $\nu$ of the Markov process} \eqref{eq-randdyn} {\sl generated by a random sequence in $\Gg$ is given by the invariant measure on the Grassmannian obtained by rotating an adequate initial condition $v^{(p)}$ with unitaries drawn from the Haar measure on the maximal compact subgroup $\Uu\subset\Gg$.} 

\vspace{.2cm}

We don't expect the RPP to hold exactly except in a model where it is imposed artificially. On the other hand, it may hold approximately in the perturbative situation described above, and its implications can also be observed in numerical experiments on particular models, provided one restricts the attention to so-called elliptic channels ({\it e.g.} \cite{RS0}). As discussed in detail in \cite{RS}, the RPP is strictly weaker than the  maximal entropy Ansatz \cite{Dor,MPK} used for the derivation of the DMPK equations. A simple toy model for which the RPP holds exactly (and thus all the below as well) is obtained by requiring $R_n$ in \eqref{eq-perturbform} to be Haar distributed on $\Uu\subset\Gg$ and then $P$ to be distributed independently according to an arbitrary measure on the Lie algebra of $\Gg$. Then it is obvious that the invariant measure $\nu$ of the Markov process \eqref{eq-randdyn} is given by the Haar measure, {\it i.e.}, the RPP holds.  In general, provided the RPP holds, the calculation of the Lyapunov spectrum by Furstenberg's formula \eqref{eq-Furstenberg} becomes
\begin{equation}
\label{eq-FurstenbergRPP}
\sum_{q=1}^p\gamma_q
\;=\;
\EE_T\;\EE_\Uu\;
\log
\left(
\|\Lambda^pT\,\Lambda^pU\,v_1\wedge\ldots\wedge v_p\|
\right)
\;.
\end{equation}
In a perturbative regime which is of interest for several applications, the evaluation of \eqref{eq-FurstenbergRPP} is now an algebraic exercise. This was actually carried out in detail for the groups $\Gg^{\mbox{\rm\tiny AIII}}$, $\Gg^{\mbox{\rm\tiny CI}}$, and $\Gg^{\mbox{\rm\tiny DIII}}$ in \cite{RS}. Here we also consider the remaining cases which involves calculating averages not only over the unitary, but also the orthogonal and symplectic group. The basic tool will be the following perturbative formula in the case where $T=Re^{\lambda P}$ with $R$ unitary.  Let $v_1,\ldots,v_p$ be orthonormal vectors in $\CM^N$, and set $\pi_p=\sum_{q=1}^pv_q v_q^*$. Then
\begin{eqnarray}
\log
\left(
\|\Lambda^p Re^{\lambda P}\,\Lambda^pU\,v_1\wedge\ldots\wedge v_p\|
\right)
&  &
\nonumber
\\
&  & 
\hspace{-5.6cm}
=\;
\frac{\lambda}{2}\;
\Tr\left(U^*Q\,U\pi_p\right)
\;+\;
\frac{\lambda^2}{4}\;
\Tr\left(U^*S\,U\pi_p\right)
%\nonumber
%\\
%& & \;\;
\;-\;
\frac{\lambda^2}{4}\;
\Tr\left(U^*Q\,U\pi_pU^*QU\pi_p\right)
\;+\;
\Oo(\lambda^3)
\;,
\label{eq-expan}
\end{eqnarray}
where
$$
Q\;=\;P+P^*\;,
\qquad
S\;=\;P^2+2\,P^*P+(P^*)^2
\;.
$$
To check this, let us begin by recalling the definition of the norm in $\Lambda^p\CM^N$:
$$
\log
\left(
\|\Lambda^p Re^{\lambda P}\,\Lambda^pU\,v_1\wedge\ldots\wedge v_p\|
\right)
\;=\;
\frac{1}{2}\;
\log
\;\mbox{\rm det}_p\,
\left(
v_i^*(e^{\lambda P}U)^*e^{\lambda P}U v_j
\right)_{i,j=1,\ldots,p}
\;.
$$
Next the identity $\log \mbox{\rm det}_p=\Tr_p\log$ and an expansion in $\lambda$ shows equality to
$$
\frac{1}{2}\;
\Tr_p\,\log
\left(\one_p\;+\;
\lambda\,\left(v_i^*U^*QU v_j
\right)_{i,j=1,\ldots,p}
\;+\;
\frac{\lambda^2}{2}\,\left(v_i^*U^*SU v_j
\right)_{i,j=1,\ldots,p}
\,+\,
\Oo(\lambda^3)
\,
\right)
\;.
$$
Now an expansion of the logarithm already proves \eqref{eq-expan}. Substituting \eqref{eq-expan} into \eqref{eq-FurstenbergRPP} one obtains a perturbative formula for the sum of the Lyapunov exponents. Taking the difference of two such sums then allows to deduce a perturbative formula for each Lyapunov exponent $\gamma_p$. This procedure and also the above proof of  \eqref{eq-expan} do not provide good error estimates on the dependence of the $\Oo(\lambda^3)$ term on the matrix size $N$. Actually, considerably improved bounds can be obtained using the Gram-Schmidt cocycle with values in the upper triangular matrices with positive diagonal \cite{RS}. Further improvements in this direction will be provided elsewhere. Here the main focus is on the calculation of the contributions up to order $\Oo(\lambda^2)$ rather than the error estimate. Let us now come to the results in each of the 10 Cartan classes. Even though the final formulas are all relatively compact, it ought to be added that the calculations leading to these formulas are quite tedious. We only provide the main intermediate steps. 

%%%%%%%%%%%%%%%%%%%%%%%%%%%%%%%%%%%%%%%%%%%%%
\subsection{Lyapunov spectrum for Class A}
\label{sec-LyapA}

As this is the first case, let us treat it with some more details. Due to Section~\ref{sec-GA}, the group is $\Gg^{\mbox{\rm\tiny A}}=\mbox{\rm GL}(N,\CM)$ with maximal compact subgroup $\Uu^{\mbox{\rm\tiny A}}=\mbox{\rm U}(N)$. As there are no specific symmetries, one may choose the unit vectors $v_q$ to be the standard basis vectors $e_q$, having only one non-vanishing entry in the $q$th component. Then, replacing \eqref{eq-expan} into \eqref{eq-FurstenbergRPP}, one sees that one needs to calculate second and fourth moments of the Haar measure on $\mbox{\rm U}(N)$. The formulas needed are listed in Lemma~\ref{lem-Umoments} of the Appendix. As $\Tr(\pi_p)=p$, one deduces with some care
$$
\sum_{q=1}^p\gamma_q^{\mbox{\rm\tiny A}}
\;=\;
\EE_P\left[
\lambda\,\frac{p\,\Tr(Q)}{2N}
\,+\,
\lambda^2\,\frac{p\,\Tr(S)}{4N}
\,-\,
\lambda^2\,
\frac{(Np-p^2)\Tr(Q)^2+(Np^2-p)\Tr(Q^2)}{4N(N^2-1)}
\right]
\;+\;\Oo(\lambda^3)
\;.
$$
Now the identity $\Tr(S)=\Tr(Q^2)$ allows to further simplify. Furthermore, taking the difference of $\sum_{q=1}^p\gamma_q^{\mbox{\rm\tiny A}}$ and $\sum_{q=1}^{p-1}\gamma_q^{\mbox{\rm\tiny A}}$, one finds
\begin{equation}
\label{eq-LyapA}
\gamma_p^{\mbox{\rm\tiny A}}
\;=\;
\lambda\,\frac{1}{2N}\,\EE_P\;\Tr(Q)
\;+\;
\lambda^2\,\frac{N+1-2p}{4(N^2-1)}
\;\EE_P\left(\Tr(Q^2)-\frac{1}{N}\,\Tr(Q)^2\right)
\;+\;\Oo(\lambda^3)
\;.
\end{equation}
The main remarkable facts about this formula are the following. To order $\lambda$, the Lyapunov spectrum is $N$-fold degenerate. However, in second order perturbation theory, this degeneracy is lifted and, furthermore, the spacings between the Lyapunov exponents are equal. This equidistance is observed in numerous numerical experiments and is here shown to be a consequence of the RPP. Finally, the Lyapunov exponent $\gamma_{[\frac{N+1}{2}]}^{\mbox{\rm\tiny A}}$ is smallest in absolute value. Here $[\,.\,]$ denotes the integer part. If $N$ is odd, then $\gamma_{[\frac{N+1}{2}]}^{\mbox{\rm\tiny A}}=0$ if $P$ is centered. However, this vanishing is lifted by higher order terms, just as a non-centered $P$ would not necessarily lead to a non-vanishing Lyapunov exponent. Hence the vanishing is not symmetry enforced, as in Classes DIII and D with odd $N$ discussed below.

%%%%%%%%%%%%%%%%%%%%%%%%%%%%%%%%%%%%%%%%%%%%%
\subsection{Lyapunov spectrum for Class AI}
\label{sec-LyapAI}

Here all matrices are real, since $\Gg^{\mbox{\rm\tiny AI}}=\mbox{\rm GL}(N,\RM)$. Thus also the initial vectors $v_1,\ldots,v_p$ should be chosen real. Thus one can again choose $v_q=e_q$. Then one is led to an analogous calculation as in Class A, but according to the RPP, the averages now have to be taken over the orthogonal group instead of the unitary group. The corresponding formulas are collected in Lemma~\ref{lem-Omoments}. After some algebra one finds
$$
\gamma_p^{\mbox{\rm\tiny AI}}
\;=\;
\lambda\,\frac{1}{2N}\,\EE_P\;\Tr(Q)
\;+\;
\lambda^2\,\frac{N+1-2p}{4(N-1)(N+2)}
\;\EE_P\left(\Tr(Q^2)-\frac{1}{N}\,\Tr(Q)^2\right)
\;+\;\Oo(\lambda^3)
\;.
$$
The same comments as in Class A apply.

%%%%%%%%%%%%%%%%%%%%%%%%%%%%%%%%%%%%%%%%%%%%%
\subsection{Lyapunov spectrum for Class AII}
\label{sec-LyapAII}

Here the matrices $T=Re^{\lambda P}$ are in the group $\Gg^{\mbox{\rm\tiny AII}}=\mbox{\rm U}^*(2N)$, and averages have to be taken over the compact symplectic group $\Uu^{\mbox{\rm\tiny AII}}=\mbox{\rm SP}(2N)$. However, some further care is needed with the choice of the $v_1,\ldots,v_p$. Indeed, the Lyapunov exponents are the scaling exponents of the singular values of $T_n\cdots T_1$, thus of the eigenvalues of the positive matrix $(T_n\cdots T_1)^*T_n\cdots T_1$ in the group $\mbox{\rm U}^*(2N)$. As explained in Section~\ref{sec-GAII}, these eigenvalues are twice degenerate, and the eigenvectors are pairs $v,I\overline{v}$. Thus also the Lyapunov spectrum is twice degenerate. If one chooses $v_1=e_1$ and this leads to the largest singular value and thus largest Lyapunov exponent in \eqref{eq-FurstenbergRPP}, then one has to choose $v_2=I e_1$ so that indeed the second Lyapunov exponent is equal to the first one, hence establishing the degeneracy just explained. Arguing similarly for the other Lyapunov exponents, we therefore set
$$
v_{2q-1}\;=\;e_q\;,
\qquad
v_{2q}\;=\;I\,\overline{v_{2q-1}}\;=\;I\,e_q\;=\;e_{q+N}\;.
$$
Then the projections $\pi_p=\sum_{q=1}^pv_qv_q^*=(\pi_p)^t$ are of the form
\begin{equation}
\label{eq-pipform}
\pi_p\;=\;
\begin{pmatrix} b & 0 \\ 0 & c \end{pmatrix}
\;,
\qquad
b\,=\,\sum_{q=1}^{j+\delta}e'_q (e'_q)^*\;,\;\;\;c\,=\,\sum_{q=1}^{j}e'_{q} (e'_{q})^*
\;,
\end{equation}
where $p=2j+\delta$ with $j\in\NM$ and $\delta\in\{0,1\}$, and $e'_q$ denotes again the standard basis vectors of $\CM^N$ (those in $\CM^{2N}$ being denoted by $e_q$). Note that both $b$ and $c$ are $N\times N$ matrices. They satisfy  $bc=cb=c$ as well as $\Tr(b)=j+\delta$ and $\Tr(c)=j$. This implies
$$
I^*\pi_{2j+\delta} I \,\pi_{2j+\delta}\;=\;\pi_{2j}
\;.
$$
Furthermore the formula $I^*Q^tIQ=I^*\overline{Q}IQ=Q^2$ is needed. With these identities at hand, a careful use of Lemma~\ref{lem-SPmoments} leads to
\begin{eqnarray*}
\sum_{q=1}^p\gamma_q^{\mbox{\rm\tiny AII}}
& = &
\EE_P\Bigl[
\lambda\,\frac{p\,\Tr(Q)}{4N}
\;+\;
\lambda^2\,\frac{p\,\Tr(S)}{8N}
\\
& & \;\;\;\;\;\;-\;
\lambda^2\,
\frac{(2Np+\delta-p^2)\Tr(Q)^2+(2Np^2-2p-2Np+2N\delta)\Tr(Q^2)}{16N(N-1)(2N+1)}
\Bigr]
\;+\;\Oo(\lambda^3)
\;,
\end{eqnarray*}
where $\delta=\frac{1}{2}(1-(-1)^p)$ as above. Now again $\Tr(S)=\Tr(Q^2)$ allows to somewhat simplify. Then taking differences a further calculation shows
$$
\gamma_p^{\mbox{\rm\tiny AII}}
\;=\;
\lambda\,\frac{1}{4N}\,\EE_P\;\Tr(Q)
\;+\;
\lambda^2\,\frac{2N+1-2p+(-1)^p}{8(N-1)(2N+1)}
\;\EE_P\left(\Tr(Q^2)-\frac{1}{2N}\,\Tr(Q)^2\right)
\;+\;\Oo(\lambda^3)
\;,
$$
where $p=1,\ldots, 2N$. Besides the comments made in Class A, let us note that this formula respects the double degeneracy of the Lyapunov spectrum resulting from Kramers' degeneracy.

%%%%%%%%%%%%%%%%%%%%%%%%%%%%%%%%%%%%%%%%%%%%%
\subsection{Lyapunov spectrum for Classes AIII}
\label{sec-LyapAIII}

The reflection symmetry of the spectrum of matrices in $\Gg^{\mbox{\rm\tiny AIII}}=\mbox{\rm U}(N,N)$ first of all implies that the Lyapunov spectrum satisfies $\gamma_p^{\mbox{\rm\tiny AIII}}=-\gamma^{\mbox{\rm\tiny AIII}}_{2N+1-p}$. Hence it is sufficient to calculate the non-negative Lyapunov exponents. In principle, one can proceed as above, but it is important to choose the vectors $v_q$ $J$-isotropic because the Lyapunov spectrum studies the scaling of the singular values of products in $\mbox{\rm U}(N,N)$, which are eigenvalues of positive matrices in $\mbox{\rm U}(N,N)$ so that their eigenvectors are always isotropic. Therefore we choose
\begin{equation}
\label{eq-UNNchoice}
v_q\;=\;\frac{1}{\sqrt{2}}\,(e_q+e_{q+N})
\;.
\end{equation}
It follows that $\pi_p=\frac{1}{2}\binom{d\;d}{d\;d}$ where $d=\sum_{q=1}^pe'_q(e'_q)^*$ is an $N\times N$ matrix built from the standard basis $e'_q$ in $\CM^N$. Now the average over $\Uu^{\mbox{\rm\tiny AIII}}=\mbox{\rm U}(N)\times\mbox{\rm U}(N)$ can be calculated with Lemma~\ref{lem-Umoments}(i). Furthermore, the relation $QJ+JQ=0$ implies that $Q=\binom{0\;\;a}{a^*\;0}$ is off-diagonal so that, in particular, $\Tr(Q)=0$. With the notations $\Pi_+=\binom{\one\;0}{0\;0}$ and $\Pi_-=\binom{0\;0}{0\;\one}$, and with $V, W$ as in \eqref{M-AIIIlabel} above,
 one gets after carrying out the averages over the quadratic terms of $U$:
\begin{eqnarray*}
\sum_{q=1}^p\gamma_q^{\mbox{\rm\tiny AIII}}
& = &
\EE_P\,\EE_V\,\EE_W\Bigl[
\frac{\lambda^2}{4}\,\frac{1}{2N}\,\big(\Tr(S\Pi_+)\Tr(\pi_p\Pi_+)\,+\,\Tr(S\Pi_-)\Tr(\pi_p\Pi_-)\big)
\\
& & 
\;\;\;\;\;\;\;\,-\,\frac{\lambda^2}{16}
\big(\Tr(V^*aWdW^*a^*Vd)+\Tr(W^*a^*VdV^*aWd)\big)
\Bigr]
\;+\;\Oo(\lambda^3)
\;,
%\label{eq-UNNinter}
\end{eqnarray*}
so that $\Tr(\pi_p\Pi_\pm)=\Tr(d)=p$ and $\Tr(S)=2\,\Tr(a^*a)=\Tr(Q^2)$ lead to
$$
\sum_{q=1}^p\gamma_q^{\mbox{\rm\tiny AIII}}
\;=\;
\frac{\lambda^2}{8}\;\left(\frac{p}{N}-\frac{p^2}{2N^2}\right)\,\EE_P\,\Tr(Q^2)
\;+\;\Oo(\lambda^3)\;.
$$
Taking differences therefore shows
\begin{equation}
\label{eq-LyapAIII}
\gamma_p^{\mbox{\rm\tiny AIII}}
\;=\;
\lambda^2\,\frac{N-p+\frac{1}{2}}{8\,N^2}
\;
\EE_P\,\Tr(Q^2)
\;+\;\Oo(\lambda^3)
\;,
\qquad
p=1,\ldots,N\;,
\end{equation}
which was already derived in \cite{RS}. One important feature of this formula is that the smallest non-negative Lyapunov exponent $\gamma_N^{\mbox{\rm\tiny AIII}}$ is strictly positive. This does not hold if one considers random products in $\mbox{\rm U}(N,M)$ with $N<M$. Indeed, then there are $M-N$ vanishing Lyapunov exponents. With some care one can transpose the above calculation to show that the remainder of the Lyapunov spectrum is again given by \eqref{eq-LyapAIII}, but in the denominator $N^2$ is replaced by $NM$.

%%%%%%%%%%%%%%%%%%%%%%%%%%%%%%%%%%%%%%%%%%%%%
\subsection{Lyapunov spectrum for Classes CI}
\label{sec-LyapCI}

In Class CI, $\Gg^{\mbox{\rm\tiny CI}}\cong\mbox{\rm SP}(2N,\RM)$ and $\Uu^{\mbox{\rm\tiny CI}}\cong\mbox{\rm U}(N)$. This was also dealt with in \cite{RS} and the result is
$$
\gamma_p^{\mbox{\rm\tiny CI}}
\;=\;
\lambda^2\,\frac{N-p+1}{8\,N(N+1)}
\;
\EE_P\,\Tr(Q^2)
\;+\;\Oo(\lambda^3)\;,
\qquad
p=1,\ldots,N\;.
$$
Then the other half of the Lyapunov spectrum is again given by  $\gamma_p^{\mbox{\rm\tiny CI}}=-\gamma^{\mbox{\rm\tiny CI}}_{2N+1-p}$. The proof is a mixture of Class AIII and DIII below.

%%%%%%%%%%%%%%%%%%%%%%%%%%%%%%%%%%%%%%%%%%%%%
\subsection{Lyapunov spectrum for Class DIII}
\label{sec-LyapDIII}

In Class DIII the matrices are in the group $\Gg^{\mbox{\rm\tiny AIII}}= \mbox{\rm U}^*(2N)\,\cap \,\mbox{\rm U}(N,N)\cong \mbox{\rm SO}^*(2N)$, so that the Lyapunov spectrum has both the double Kramers' degeneracy and the symmetry around $0$. For odd $N$, this implies that there must be a twice degenerate vanishing Lyapunov exponent. This vanishing is symmetry enforced and thus must also hold to all orders of perturbation theory. The implications of a vanishing Lyapunov exponent are discussed in Section~\ref{sec-deloc}. Now let us come to the perturbative calculation of the Lyapunov spectrum.  For the choice of the orthonormal basis $v_q$, two criteria have to be satisfied. First of all, the eigenvectors have to be $J$-isotropic and pairwise $J$-orthogonal for different eigenvalues as in the case of $\Gg^{\mbox{\rm\tiny AIII}}$, and second of all, they should be chosen in pairs as in the case of $\Gg^{\mbox{\rm\tiny AII}}$. Therefore, we set
\begin{equation}
\label{eq-choiceDIII}
v_{2q-1}\;=\;\frac{1}{\sqrt{2}}\,(e_q+e_{q+N})\;,
\qquad
v_{2q}\;=\;I\,\overline{v_{2q-1}}\;=\;\frac{1}{\sqrt{2}}\,(e_{q+N}-e_{q})\;.
\end{equation}
Note that both of these vectors are $J$-isotropic, but they are not $J$-orthogonal (which is not required because they correspond to eigenvectors of the same eigenvalue). However, each is $J$-orthogonal to all other basis vectors except for its partner. Then $\pi_p=\sum_{q=1}^pv_q (v_q)^*$ is precisely the same as in \eqref{eq-pipform}. Now the average over the group $\Uu^{\mbox{\rm\tiny DIII}}\cong \mbox{\rm U}(N)$ in \eqref{eq-FurstenbergRPP} is carried out using the formulas of Lemma~\ref{lem-UCImoments}. Note that $QJ+JQ=0$ and $I^*\overline{Q}I=Q$ imply that $Q=Q^*=\binom{0\;\;a}{a^*\;0}$ is off-diagonal with $a^t=-a$, so that Lemma~\ref{lem-UCImoments}(iii) applies. Also $\Tr(Q)=0$. Furthermore, $\pi_p$ is also of the form required in Lemma~\ref{lem-UCImoments}(ii) and (iii).  With the notation $\Pi_\pm$ as in Lemma~\ref{lem-UCImoments}, and writing again $p=2j+\delta$ with $j\in\NM$ and $\delta\in\{0,1\}$, 
$$
\sum_{q=1}^p\gamma_q^{\mbox{\rm\tiny DIII}}
\;=\;
\frac{\lambda^2}{4}\,
\EE_P
\left[
\frac{j+\delta}{N}\,\Tr(\Pi_+S)+\frac{j}{N}\,\Tr(\Pi_-S)
-
\frac{(j+\delta)j-j}{N(N-1)}\;\Tr(Q^2)
\right]
\;+\;\Oo(\lambda^3)
\;.
$$
As one also has $I^*\overline{S}I=S$ and $I^*\Pi_+I=\Pi_-$, it follows that $\Tr(\Pi_-S)=\Tr(\Pi_+S)=\frac{1}{2}\Tr(Q^2)$. Thus
$$
\sum_{q=1}^p\gamma_q^{\mbox{\rm\tiny DIII}}
\;=\;
\frac{\lambda^2}{16}\;
\left[\,
\frac{2\,p}{N}\;-\;
\frac{p^2-2\,p+\delta}{N(N-1)}\,
\right]\;\EE_P\,\Tr(Q^2)
\;+\;\Oo(\lambda^3)
\;.
$$
From this one deduces
$$
\gamma_{p}^{\mbox{\rm\tiny DIII}}
\;=\;
\lambda^2\,\frac{N-p+\frac{1}{2}+\frac{1}{2}\,(-1)^p}{8\,N(N-1)}
\;
\EE_P\,\Tr(Q^2)
\;+\;\Oo(\lambda^3)
\;,
\qquad
p=1,\ldots,N\;.
$$
Again the other half of the Lyapunov spectrum is given by $\gamma_p^{\mbox{\rm\tiny DIII}}=-\gamma^{\mbox{\rm\tiny DIII}}_{2N+1-p}$. As already pointed out above, the case of even $N$ was already treated in \cite{RS}.

\vspace{.2cm}

It is remarkable that the smallest non-negative Lyapunov exponents  $\gamma_N^{\mbox{\rm\tiny AIII}}$, $\gamma_N^{\mbox{\rm\tiny CI}}$ and $\gamma_N^{\mbox{\rm\tiny DIII}}$ for even $N$ are strictly positive and satisfy  $2\,\gamma_N^{\mbox{\rm\tiny AIII}}=\gamma_N^{\mbox{\rm\tiny CI}}+\Oo(N^{-1})$ and $2\,\gamma_N^{\mbox{\rm\tiny AIII}}=\gamma_N^{\mbox{\rm\tiny DIII}}+\Oo(N^{-1})$, the latter again in the case of even $N$. This reflects the change of the inverse localization length under breaking of time-reversal symmetry, see the discussion in \cite{RS}. 

%%%%%%%%%%%%%%%%%%%%%%%%%%%%%%%%%%%%%%%%%%%%%
\subsection{Lyapunov spectrum for Class BDI}
\label{sec-LyapBDI}

Because the spectrum of a self-adjoint matrix in $\Gg^{\mbox{\rm\tiny BDI}}=\mbox{\rm O}(N,N)$ comes in pairs $\lambda,\lambda^{-1}$, the Lyapunov spectrum always satisfies $\gamma_p^{\mbox{\rm\tiny BDI}}=-\gamma^{\mbox{\rm\tiny BDI}}_{2N+1-p}$. Hence it is again sufficient to calculate the first half of the spectrum.  As $\mbox{\rm O}(N,N)\subset\mbox{\rm U}(N,N)$, one has to choose the vectors $v_q$ as in \eqref{eq-UNNchoice}. These vectors are already chosen to be real, which would be the second requirement in the present case. Now both $\pi_p$ and $Q$ are as in Class AIII, $\pi_p=\frac{1}{2}\binom{d\;d}{d\;d}$ where $d=\sum_{q=1}^pe'_q(e'_q)^*$ and $Q=\binom{0\;\;a}{a^*\;0}$. The only supplementary property is that $a$ is real.  Also the calculation of the Lyapunov spectrum remains the same as in Class AIII, except that the average is now taken w.r.t. the Haar measure on $\mbox{\rm O}(N)$ instead of $\mbox{\rm U}(N)$. As only the second moments enter and they are the same by Lemma~\ref{lem-Omoments}(i) and Lemma~\ref{lem-Umoments}(i), also the final result of the calculation is the same:
$$
\gamma_p^{\mbox{\rm\tiny BDI}}
\;=\;
\lambda^2\,\frac{N-p+\frac{1}{2}}{8\,N^2}
\;
\EE_P\,\Tr(Q^2)
\;+\;\Oo(\lambda^3)
\;,
\qquad
p=1,\ldots,N\;.
$$
If one considers the case of $\mbox{\rm O}(N,M)$, the same comments as at the end of Section~\ref{sec-LyapAII} apply.

%%%%%%%%%%%%%%%%%%%%%%%%%%%%%%%%%%%%%%%%%%%%%
\subsection{Lyapunov spectrum for Class CII}
\label{sec-LyapCII}

As already stressed in Section~\ref{sec-GCII}, elements of both groups $\Gg^{\mbox{\rm\tiny CII}}$ and $\Gg^{\mbox{\rm\tiny DIII}}$ have Kramers' degeneracy and reflection symmetry, so that the Lyapunov spectrum is twice degenerate and symmetric around $0$. However, the symmetry leading to Kramers' degeneracy is different and therefore instead of \eqref{eq-choiceDIII} one may set for $q=1,\dots,N$
$$
v_{2q-1}\;=\;\frac{1}{\sqrt{2}}\,(e_q+e_{q+2N})\;,
\qquad
v_{2q}\;=\;J\otimes I\,\overline{v_{2q-1}}\;=\;\frac{1}{\sqrt{2}}\,(e_{q+N}-e_{q+3N})\;.
$$ 
From this one deduces 
$$
\pi_p
\;=\;
\begin{pmatrix}
e & f \\ f & e
\end{pmatrix}
\;,
\qquad
e\,=\,
\frac{1}{2}
\begin{pmatrix}
b & 0  \\
0 & c \\
\end{pmatrix}
\;,
\;\;\;f\,=\,
\frac{1}{2}
\begin{pmatrix}
b & 0  \\
0 & -c \\
\end{pmatrix}
\;,
$$
with $N\times N$ matrices $b$ and $c$ defined in \eqref{eq-pipform}. Thus $IfIf=\frac{1}{4} \binom{c\;0}{0\;c}$. In particular, for $p=2j+\delta$ one has $2\,\Tr(e)=p$  and $2\,\Tr(IfIf)=j$. Furthermore, $Q=\binom{0\;\;a}{a^*\;0}$ with $I\overline{a}I=a$ (note the sign which means that $a$ is not a quaternion matrix, but $\imath\,a$ is).  Now from Lemma~\ref{lem-UCIImoments} it follows
that
\begin{eqnarray*}
\sum_{q=1}^p\gamma_q^{\mbox{\rm\tiny CII}}
& = &
\frac{\lambda^2}{4}\,\EE_P
\Big[\,
\frac{1}{2N}\,\bigl(\Tr(\Pi_+S)\Tr(\Pi_+\pi_p)+\Tr(\Pi_-S)\Tr(\Pi_-\pi_p)\bigr)
\\
& &
\hspace{1.5cm}
 -\,
\frac{1}{4N^2}\,\Tr(Q^2)\bigl(\Tr(e)^2+\Tr(I^*fIf)\bigr)
\Big]
\;+\;\Oo(\lambda^3)
\;,
\end{eqnarray*}
Replacing thus shows
$$
\sum_{q=1}^p\gamma_q^{\mbox{\rm\tiny CII}}
\;=\;
\frac{\lambda^2}{4}\,
\left[\,
\frac{p}{4N}-
\frac{1}{4N^2}\Big( \frac{p^2}{4}-\frac{j}{2}\Big) \,
\right]\,\EE_P\,\Tr(Q^2)
\;+\;\Oo(\lambda^3)
\;.
$$
Therefore
$$
\gamma_p^{\mbox{\rm\tiny CII}}
\;=\;
\frac{\lambda^2}{8}\;
\frac{2N-p+1+\frac{1}{2}\,(-1)^p}{4N^2}\;\EE_P\,\Tr(Q^2)
\;+\;\Oo(\lambda^3)
\;,
\qquad
p=1,\ldots,2N\;.
$$

%%%%%%%%%%%%%%%%%%%%%%%%%%%%%%%%%%%%%%%%%%%%%
\subsection{Lyapunov spectrum for Class D}
\label{sec-LyapD}

The reflection symmetry of the spectrum of a positive matrix in $\Gg^{\mbox{\rm\tiny D}}=\mbox{\rm O}(N,\CM)$ discussed in Section~\ref{sec-GD} implies again that the Lyapunov spectrum always satisfies $\gamma_p^{\mbox{\rm\tiny D}}=-\gamma^{\mbox{\rm\tiny D}}_{N+1-p}$. Therefore, if $N$ is odd, there is one vanishing Lyapunov exponent $\gamma_{\frac{N+1}{2}}^{\mbox{\rm\tiny D}}=0$. This vanishing is hence symmetry enforced. For the calculation of the Lyapunov spectrum, one starts again from \eqref{eq-FurstenbergRPP} and \eqref{eq-expan}. According to the RPP the average now has to be calculated over the group $\Uu^{\mbox{\rm\tiny D}}=\mbox{\rm O}(N)$, thus depends on the formulas in Lemma~\ref{lem-Omoments}. Furthermore, one uses that $Q=Q^*\in\mbox{\rm so}(N,\CM)$ satisfies $\Tr(Q)=0$ and $Q=-Q^t=-\overline{Q}$, hence $\Tr(Q^tQ)=-\Tr(Q^2)$. From this and $\Tr(S)=\Tr(Q^2)$ one deduces after some algebra
$$
\sum_{q=1}^p\gamma_q^{\mbox{\rm\tiny D}}
\;=\;
\frac{\lambda^2}{4}\,
\left[\,
\frac{p}{N}-
\frac{p^2-p}{N(N-1)}\,
\right]\,\EE_P\,\Tr(Q^2)
\;-\;
\frac{\lambda^2}{4}\,
\frac{Np-p^2}{N(N-1)(N+2)}
\,\EE_P\,\Tr(Q)^2
\;+\;\Oo(\lambda^3)
\;,
$$
which implies 
$$
\gamma_{p}^{\mbox{\rm\tiny D}}
\;=\;
\frac{\lambda^2}{4}\;
\frac{N+1-2p}{N(N-1)}
\;\EE_P
\left[
\Tr(Q^2)
-
\frac{1}{N+2}\,\Tr(Q)^2
\right]
\;+\;\Oo(\lambda^3)
\;,
\qquad
p=1,\ldots,N\;.
$$
Hence in Class D there is never a term of order $\Oo(\lambda)$, and the perturbative formula indeed respects $\gamma_{\frac{N+1}{2}}^{\mbox{\rm\tiny D}}=0$.

%%%%%%%%%%%%%%%%%%%%%%%%%%%%%%%%%%%%%%%%%%%%%
\subsection{Lyapunov spectrum for Class C}
\label{sec-LyapC}

Similar as in Class D the Lyapunov spectrum satisfies $\gamma_p^{\mbox{\rm\tiny C}}=-\gamma^{\mbox{\rm\tiny C}}_{2N+1-p}$, but as $2N$ is even, this symmetry enforces no vanishing of a Lyapunov exponent. As preparation for the calculation of the Lyapunov spectrum, let us note that $Q=Q^*\in\mbox{\rm sp}(2N,\CM)$ satisfies $Q=-I^*Q^tI=-I^*\overline{Q}I$ and thus $\Tr(Q)=0$ and $\Tr(I^*Q^tIQ)=-\Tr(Q^2)$. By the RPP, the average in \eqref{eq-FurstenbergRPP} is calculated over the group $\Uu^{\mbox{\rm\tiny C}}=\mbox{\rm SP}(2N)$, based on Lemma~\ref{lem-SPmoments}. One finds, first for $p\leq N$,
$$
\sum_{q=1}^p\gamma_q^{\mbox{\rm\tiny C}}
\;=\;
\frac{\lambda^2}{4}\,
\EE_P\left[\,
\frac{p}{2N}\,\Tr(S)-
\frac{p^2}{2N(2N+1)}\,\Tr(Q^2)\,
\right]
\;+\;\Oo(\lambda^3)
\;,
$$
and therefore
$$
\gamma_{p}^{\mbox{\rm\tiny C}}
\;=\;
\frac{\lambda^2}{4}\;
\frac{N+1-p}{N(2N+1)}
\;\EE_P
\,\Tr(Q^2)
\;+\;\Oo(\lambda^3)
\;,
\qquad
p=1,\ldots,N\;.
$$
The second (negative) half of the Lyapunov spectrum is again given by reflection.

%%%%%%%%%%%%%%%%%%%%%%%%%%%%%%%%%%%%%%%%%%%%%%%%%%%%%%%%%%%%%%%%%%%%
\section{Symmetries of one-particle Hamiltonians}
\label{sec-sym}

This section briefly reviews the ten-fold way of Altland and Zirnbauer \cite{AZ}, albeit with a classification of the symmetries in the spirit of \cite{SRFL,RSFL}. In this classification the one-particle (first quantized) Hamiltonian $H$ acting on a complex Hilbert space can have three basic symmetries, the time reversal symmetry (TRS), the particle-hole symmetry (PHS) and the sublattice or chiral symmetry (SLS). The first two of these symmetries can be either even or odd. These symmetries are implemented by real unitaries $J$, $I$, $K$, $L$ and $M$ in the following manner:
\begin{eqnarray}
\label{eq-evenTRS}
\mbox{\rm (even TRS)} & &
J^*\,\overline{H}\,J\;=\;H\;,\;\;\;\;\;
\qquad
\;\;\;J^2\;=\;\one\;,
\\
\label{eq-oddTRS}
\mbox{\rm (odd TRS)} & &
I^*\,\overline{H}\,I\;=\;H\;,\;\;\;\;\;
\qquad
\;\;\;\;I^2\;=\;-\,\one\;,
\\
\label{eq-evenPHS}
\mbox{\rm (even PHS)} & &
K^*\,\overline{H}\,K\;=\;-\,H\;,\;\;
\qquad
\,K^2\;=\;\one\;,
\\
\label{eq-oddPHS}
\mbox{\rm (odd PHS)} & &
L^*\,\overline{H}\,L\;=\;-\,H\;,\;\;
\qquad
\;\;L^2\;=\;-\,\one\;,
\\
\label{eq-SLS}
\mbox{\rm (SLS)} & &
M^*\,{H}\,M\;=\;-\,H\;,
\qquad
\;M^2\;=\;\pm\,\one\;.
\end{eqnarray}
The notational choice of letters $J$, $I$, {\it etc.}, is not yet connected to the symmetries as defined in \eqref{eq-symmdef}, but will in some cases be so in our examples later on. For instance, it is possible to have $J=\one$ so that the even PHS just reads $\overline{H}=H$. 

\vspace{.2cm}

Let us add a few comments. A given Hamiltonian $H$ can have only one TRS and one PHS symmetry. If one TRS and one PHS is present, then one also has a SLS symmetry. For example, if \eqref{eq-oddTRS} and \eqref{eq-evenPHS} hold, then also \eqref{eq-SLS} with $M=K^*I$. This also imposes the sign of the SLS which is why one does not keep track of it. The only case in which this sign may seem interesting is the case where there is only SLS. But in this situation the reality of $M$ is irrelevant and then $M$ may be changed to $\imath\,M$ which alters the sign. Counting all possibilities to combine the three symmetries, one obtains 10 classes which are listed in Table~1 ($1$ with no symmetry, $5$ with one symmetry and $4$ with three symmetries). The table also contains the Cartan classification associated to each such $H$ Class by Altland and Zirnbauer \cite{AZ} in the following manner. Each Cartan class is specified by an involution on a compact Lie algebra. The $-1$ eigenspace of the involution multiplied with $\imath$ then produces a set of self-adjoint matrices which forms the corresponding $H$ Class of Hamiltonians. See also \cite{HHZ,SRFL,RSFL,SCR,AK} for further insights on this connection.

%\begin{table*}[t]
\begin{table*}
%\begin{table}
\label{table1}
\begin{center}
\begin{tabular}{|c||c|c|c||c|c|c|c|c|}
\hline
$H$ Class & TRS&PHS &SLS & $\Gg$ Class & $\Gg$    &  $\Uu\,\cong$ & $\;\mu\;$
\\\hline\hline
A    &$0$ &$0$&$0$& AIII & $ \mbox{\rm U}(N,M)$&$ \mbox{\rm U}(N)\!\times\! \mbox{\rm U}(M)$ & 1
\\
AI &$+1$&$0$&$0$&CI &  $ \mbox{\rm SP}(2N,\RM)$& $\mbox{\rm U}(N)$ & 1
%\\
%AII &$-1$&$0$&$0$& DIII&  $2N$ & $ \mbox{\rm SO}^*(4N) $&$ \mbox{\rm U}(2N)$ & 2
\\
AII &$-1$&$0$&$0$& DIII&  $ \mbox{\rm SO}^*(2N) $&$ \mbox{\rm U}(N)$
%$\!\times\! \mbox{\rm U}(N)$ 
& 2
\\
\hline\hline
AIII &$0$&$0$&$ 1$&A & ${\mbox{\rm GL}(N,\CM)}$ &$ \mbox{\rm U}(N)$ & 1(2)
\\
BDI &$+1$&$+1$&$1$&AI  &$ \mbox{\rm GL}(N,\RM)$& $ \mbox{\rm O}(N)$ & 1(2)
\\
CII  &$-1$&$-1$&$1$&AII&  $ \mbox{\rm U}^*(2N)$& $ \mbox{\rm SP}(2N)$ & 2(4)
\\
\hline \hline
D &$0$ &$+1$&$0$ & BDI & $\mbox{\rm O}(N,M)$ &$\mbox{\rm O}(N)\!\times \!\mbox{\rm O}(M)$ & 1
\\
C &$0$ &$-1$&$0$ & CII&   $\mbox{\rm SP}(2N,2M)$ & $ \mbox{\rm SP}(2N)\!\times\! \mbox{\rm SP}(2M) $ & 1
\\
DIII &$-1$&$+1$&$1$ &D & $ \mbox{\rm O}(N,\CM)$ & $ \mbox{\rm O}(N)$ & 2
\\
CI &$+1$&$-1$&$1$ & C&   $ \mbox{\rm SP}(2N,\CM) $& 
$ \mbox{\rm SP}(2N)$  & 2
\\ \hline
\end{tabular}
\caption{
Table with symmetries taken from \cite{TBFM,SRFL} together with the transfer matrix group $\Gg$ for quasi-one-dimensional operators $H$ and its maximal compact subgroup $\Uu$. Furthermore, $\mu$ is the symmetry imposed multiplicity of the Lyapunov spectrum given by the spectral multiplicity of self-adjoint elements of $\Gg$. The numbers in parenthesis give the multiplicity to lowest order in perturbation theory for centered perturbations.
}
\end{center}
\end{table*}
%\end{table}

%%%%%%%%%%%%%%%%%%%%%%%%%%%%%%%%%%%%%%%%%%%%%%%%
\section{Quasi-one-dimensional model operators}
\label{sec-model}

In this section we construct for every $H$ Class in Table~1 a quasi-one-dimensional model operator of Dirac type. If the symmetries are organized using the invariance under representations of the real and complex Clifford algebras \cite{HHZ,Kit,SCR,AK}, then there is a nice constructive way to obtain these model operators. Even though this might be quite illuminating, it is not the focus of the present paper where we only restrict ourselves to providing one example for each given class, similar as in \cite{BMSA,BFGM,TBFM}. Our model operator of Dirac type will have a random perturbation and the study of its fundamental solutions (transfer matrices) then leads to one of the classes of random products described in Section~\ref{sec-Lyapspec}. The model operators act on the Hilbert space $L^2(\RM,\CM^{2N})$ or, if necessary, on $L^2(\RM,\CM^{2N})\otimes\CM^2$, and are of one of the three following forms:
\begin{equation}
\label{eq-Ham}
H_J\;=\;J\,\imath \partial \;- \;\lambda\,\Vv\;,
\qquad
H_I\;=\;I\,\partial \;- \;\lambda\,\Vv\;,
\qquad
H_K\;=\;K\,\imath \partial \;- \;\lambda\,\Vv\;,
\end{equation}
Here $J$, $I$ and $K$ are the matrices given in \eqref{eq-symmdef} acting on $\CM^{2N}$ and the supplementary $\CM^2$ allows to implement a further symmetry (this is only necessary for $H$ Classes C, CI and CII because an odd PHS cannot be implemented in any of the three model Hamiltonians). Furthermore, $\partial$ is the derivative on $L^2(\RM,\CM^{2N})$ and $\Vv=\Vv^*$ is a matrix-valued potential. It could be a matrix-valued function, but for sake of concreteness let us rather only consider a potential given by a sum of Dirac peaks:
\begin{equation}
\label{eq-Pot}
\Vv(x)\;=\;\sum_{n\in\ZM}\Vv_n\,\delta(x-n)\;.
\end{equation}
The $\Vv_n$ are independent and identically distributed random matrices. The formal definition of the singular potential is via boundary conditions, namely either $\psi(n+)=e^{\lambda\,\imath J \Vv_n}\psi(n-)$ or $\psi(n+)=e^{\lambda\,I \Vv_n}\psi(n-)$ or $\psi(n+)=e^{\lambda\,\imath\,K\, \Vv_n}\psi(n-)$, depending on which model in \eqref{eq-Ham} is considered. These formulas combined with Weyl-Titchmarch theory allow to show that $H_J$, $H_I$ and $H_K$  are well-defined self-adjoint operators (see \cite{SS} for details). 

\vspace{.2cm}

In $H$ Classes A,C and D it is possible to consider $H_G=G\,\imath \partial - \lambda\,\Vv$ acting on $L^2(\RM,\CM^{N+M})$ where $G$ is defined in \eqref{eq-UNM} and $N<M$. This models a system with a different number of left and right movers and gives an effective description of chiral edge states of quantum Hall systems (conventional, spin and thermal quantum Hall effect respectively), which is reflected by the fact that the Lyapunov spectrum contains at least $M-N$ vanishing Lyapunov exponents. These systems will be discussed briefly in Section~\ref{sec-deloc}.

\vspace{.2cm}

For each real energy $E$, the fundamental solutions $T^E(x,y)$ of the Schr\"odinger equation $HT^E(x,y)=ET^E(x,y)$ which are right continuous in both $x$ and $y$  and satisfy the initial condition $T^E(y,y)=\one$ are also called the transfer matrices of the system. By the above, they are given by
\begin{equation}
\label{eq-Tsolution}
T^E(n-,1-)\;=\;T_n\cdots T_1\;,
\end{equation}
where the matrices $T_n$ for three cases in \eqref{eq-Ham}  are respectively given by 
$$
T_n\;=\;e^{\imath\,E\,J}e^{\lambda\,\imath\,J\,\Vv_n}\;,
\qquad
T_n\;=\;e^{E\,I}e^{\lambda\,I\,\Vv_n}\;,
\qquad
T_n\;=\;e^{\imath\,E\,K}e^{\lambda\,\imath\,K\,\Vv_n}\;.
$$
These matrices are of the form \eqref{eq-perturbform} if one sets $R_n=e^{\imath\,E\,J}$ and $P_n=\imath\,J\,\Vv_n$ in the first case, and correspondingly in the other cases. Now by construction, the transfer matrices are in the following three groups:
\begin{equation}
\label{eq-transfergroup}
\Tt_J\;=\;\mbox{\rm U}(N,N)\;,
\qquad
\Tt_I\;=\;C^*\,\mbox{\rm U}(N,N)\,C\;,
\qquad
\Tt_K\;=\;C\,\mbox{\rm U}(N,N)\,C^*
\;,
\end{equation}
where $C$ is the Cayley transform defined in \eqref{eq-Cayley}. For $H$ Class C, CI and CII one has to replace $\mbox{\rm U}(N,N)$ by $\mbox{\rm U}(2N,2N)$ due to the supplementary $\CM^2$. Thus it looks like the transfer matrices are always in the Cartan Class AIII, but actually the game in the following subsections is to show how the symmetries of the Hamiltonian (TRS, PHS, SLS) lead to subgroups of $\Tt_J$, $\Tt_I$ or $\Tt_K$ which then actually cover all Cartan classes. Table~1 collects the results of this correspondence. 

%%%%%%%%%%%%%%%%%%%%%%%%%%%%%%%%%%%%%%%%%%%%%%%%%%%%%%%%%%%%%%%%%%%%
\subsection{Standard unitary class ($H$ Class A)}
\label{sec-A}

Let us choose the model $H_J$. As there are no further symmetries, the transfer matrix group $\Tt_J^{\mbox{\rm\tiny A}}$ is equal to $\Gg^{\mbox{\rm\tiny AIII}}=\mbox{\rm U}(N,N)$ and therefore the results of Section~\ref{sec-LyapAIII} can be directly applied to study the associated Lyapunov spectrum.

%%%%%%%%%%%%%%%%%%%%%%%%%%%%%%%%%%%%%%%%%%%%%%%%%%%%%%%%%%%%%%%%%%%%
\subsection{Standard orthogonal class ($H$ Class AI)}
\label{sec-AI}

Let us again choose the model $H_J$ and then implement the even TRS by $K^*\overline{H_J}K=H_J$. Then the equalities
$$
K^*\,\overline{T_n}\,K
\;=\;
K^*\,e^{-\imath\,E\,J}\,e^{-\lambda\,\imath\,J\,\overline{\Vv_n}}
\,K
\;=\;
e^{\imath\,E\,J}\,e^{\lambda\,\imath\,J\,K^*\overline{\Vv_n}K}
\;=\;
T_n
\;
$$
show that $\Tt_J^{\mbox{\rm\tiny AI}}=\Gg^{\mbox{\rm\tiny CI}}=C\,\mbox{\rm SP}(2N,\RM)\,C^*$, and Section~\ref{sec-LyapCI} applies. Alternatively, one can choose $H_I$ and then implement the even TRS by $\overline{H_I}=H_I$. Then $\Tt_I^{\mbox{\rm\tiny AI}}=\mbox{\rm SP}(2N,\RM)$.

%%%%%%%%%%%%%%%%%%%%%%%%%%%%%%%%%%%%%%%%%%%%%%%%%%%%%%%%%%%%%%%%%%%%
\subsection{Standard symplectic class ($H$ Class AII)}
\label{sec-AII}

Starting again from the model $H_J$, the odd TRS is now $I^*\overline{H_J}I=H_J$. By a similar calculation as above one finds that $\Tt_J^{\mbox{\rm\tiny AII}}=\Gg^{\mbox{\rm\tiny DIII}}=C\,\mbox{\rm SO}^*(2N)\,C^*$. Thus Section~\ref{sec-LyapDIII} applies.

%%%%%%%%%%%%%%%%%%%%%%%%%%%%%%%%%%%%%%%%%%%%%%%%%%%%%%%%%%%%%%%%%%%%
\subsection{Chiral unitary class ($H$ Class AIII)}
\label{sec-AIII}

When there is a chiral symmetry, it is good to work with the model operator $H_I$ and implement the SLS by $J^*H_IJ=-H_I$. The reason is that this implies $J^*T_nJ=T_n$, which combined with $T_n^*IT_n=I$ shows that $T_n$ is in the group
\begin{equation}
\label{eq-GAIII}
\Tt_I^{\mbox{\rm\tiny AIII}}\;=\;
\left\{
\left.
\left(\begin{array}{cc} A & 0 \\ 0 & (A^{-1})^* \end{array}\right)
\,\right|\,
A\in \mbox{\rm GL}(N,\CM)\;
\right\}
\;.
\end{equation}
Therefore $\Tt_I^{\mbox{\rm\tiny AIII}}\!\cong \Gg^{\mbox{\rm\tiny A}}$ and the results of Section~\ref{sec-LyapA} for Class A apply. The formula in Section~\ref{sec-LyapA} gives the Lyapunov spectrum of the upper component in $\Tt_I^{\mbox{\rm\tiny AIII}}$. The exponents of the lower component are then given exactly by the negative of the upper component. In this manner one recovers the symmetry of the Lyapunov spectrum which has to hold for the group $\Tt_I^{\mbox{\rm\tiny AIII}}\subset \mbox{\rm U}(N,N)$. It is worth mentioning the following implication of the comments at the end of Section~\ref{sec-LyapA}. If the potentials $\Vv_n$ are centered with a distribution that is also even, then already the Lyapunov spectrum of the upper component given in \eqref{eq-LyapA} is symmetric around $0$ up to terms of order $\Oo(\lambda^4)$. Therefore, the Lyapunov spectrum in $\Tt_I^{\mbox{\rm\tiny AIII}}$ has a double degeneracy up to terms of order $\Oo(\lambda^4)$. If, moreover, $N$ is odd, then there are two Lyapunov exponents that vanish to second order perturbation theory and the first non-vanishing contribution appears only at order $\Oo(\lambda^4)$. This leads to a very large, but finite localization length in such systems (with small randomness). The same comments apply to the other two chiral classes.

%%%%%%%%%%%%%%%%%%%%%%%%%%%%%%%%%%%%%%%%%%%%%%%%%%%%%%%%%%%%%%%%%%%%
\subsection{Chiral orthogonal class ($H$ Class BDI)}

Everything of the last section transposes, but, moreover, one has the  even TRS $\overline{H_I}=H_I$. Combined one also obtains an even PHS $J^*\overline{H_I}J=-H_I$. The transfer matrices are now, on top of being in $\Tt_I^{\mbox{\rm\tiny AIII}}$, real as in Section~\ref{sec-AI} and therefore
$$
\Tt_I^{\mbox{\rm\tiny BDI}}\;=\;
\left\{
\left.
\left(\begin{array}{cc} A & 0 \\ 0 & (A^{-1})^* \end{array}\right)
\,\right|\,
A\in \mbox{\rm GL}(N,\RM)\;
\right\}
\;\cong\;
\Gg^{\mbox{\rm\tiny AI}}
\;.
$$
Thus the results of Section~\ref{sec-LyapAI} apply, as well as the comments made in Section~\ref{sec-AIII}. 

%%%%%%%%%%%%%%%%%%%%%%%%%%%%%%%%%%%%%%%%%%%%%%%%%%%%%%%%%%%%%%%%%%%%
\subsection{Chiral symplectic class ($H$ Class CII)}
\label{sec-CII}

As in the other two chiral classes, the Hamiltonian is $H_I$ with SLS $J^*H_IJ=-H_I$, but now it acts on $L^2(\RM,\CM^{2N})\otimes\CM^2$ and the supplementary $\CM^2$ is used to implement the odd TRS by $\one\otimes I^*\,\overline{H_I}\,\one\otimes I=H_I$ (note that here the $I\otimes \one$ in the definition of $H_I$ and $\one\otimes I$ in the odd TRS are in different gradings). As usual the two symmetries together imply another one, namely an odd PHS $J\otimes I^*\,\overline{H_I}\,J\otimes I=-H_I$. One deduces that the transfer matrices satisfy $\one\otimes I^*\,\overline{T_n}\,\one\otimes I=T_n$ so that
$$
\Tt_I^{\mbox{\rm\tiny CII}}\;=\;
\left\{
\left.
\left(\begin{array}{cc} A & 0 \\ 0 & (A^{-1})^* \end{array}\right)
\,\right|\,
A\in \mbox{\rm U}^*(2N)\;
\right\}
\;\cong\;
\Gg^{\mbox{\rm\tiny AII}}
\;.
$$
Thus the results of Section~\ref{sec-LyapAII} apply, which lead in particular to Kramers' degeneracy, which leads even to a fourfold degeneracy up to order $\Oo(\lambda^4)$ if the distribution of $\Vv_n$ is even.

%%%%%%%%%%%%%%%%%%%%%%%%%%%%%%%%%%%%%%%%%%%%%%%%%%%%%%%%%%%%%%%%%%%%
\subsection{Even PHS without TRS ($H$ Class D)}
\label{sec-frameClassD}

Here let us choose the model $H_J$ and implement the even PHS simply by $\overline{H_J}=-H_J$. This implies for the transfer matrices that $\overline{T_n}=T_n$, which leads to the following subgroup of $\Tt_J^{\mbox{\rm\tiny A}}$:
$$
\Tt_J^{\mbox{\rm\tiny D}}\;=\;
\left\{
T\in\mbox{\rm GL}(2N,\CM)\,\left|\,
T^*J \,T\,=\,J\;,\;\overline{T}=T\,
\right.
\right\}
\;=\;
\mbox{\rm O}(N, N)
\;=\;
\Gg^{\mbox{\rm\tiny BDI}}
\;.
$$
Thus the formulas of Section~\ref{sec-LyapBDI} apply.

%%%%%%%%%%%%%%%%%%%%%%%%%%%%%%%%%%%%%%%%%%%%%%%%%%%%%%%%%%%%%%%%%%%%
\subsection{Odd PHS class without TRS ($H$ Class C)}

The odd PHS cannot be implemented in any of the three model Hamiltonians, unless they are tensorized with an extra $\CM^2$. Having done that, let us choose the model $H_J$. Then an odd PHS is given by $J\otimes I^*\,\overline{H_J}\,J\otimes I=-H_J$. From this one deduces $J\otimes I^*\,\overline{T_n}\,J\otimes I=T_n$ so that the transfer matrix group is
$$
\Tt_J^{\mbox{\rm\tiny C}}\;=\;
\left\{
T\in\mbox{\rm GL}(4N,\CM)\,\left|\,
T^*J\otimes\one \,T\,=\,J\otimes\one\;,\;J\otimes I^*\,\overline{T}\,J\otimes I\,=\,T\,
\right.
\right\}
\;=\;
\Gg^{\mbox{\rm\tiny CII}}
\;.
$$
Thus the Lyapunov spectrum can be calculated as in Section~\ref{sec-LyapCII}.

%%%%%%%%%%%%%%%%%%%%%%%%%%%%%%%%%%%%%%%%%%%%%%%%%%%%%%%%%%%%%%%%%%%%
\subsection{Even PHS with odd TRS ($H$ Class DIII)}

In Class DIII let us use the model operator $H_K$ and implement the odd TRS by $I^*\overline{H_K}I=H_K$ and the even PHS by $K^*\overline{H_K}K=-H_K$. Then one readily checks that the transfer matrices are in the following group:
$$
\Tt_K^{\mbox{\rm\tiny DIII}}\;=\;
\left\{
T\in\mbox{\rm GL}(2N,\CM)\,\left|\,
T^*K \,T\,=\,K\;,\; I^*\overline{T}\,I\,=\,T\,,\;\;K^*\overline{T}\,K\,=\,T
\right.
\right\}
\;.
$$
Writing out the latter relations implies (first deduce $J^*TJ=T$ and  then use it)
$$
\Tt_K^{\mbox{\rm\tiny DIII}}\;=\;
\left\{
\left.
\left(\begin{array}{cc} A & 0 \\ 0 & (A^{-1})^* \end{array}\right)
\,\right|\,
A\in \mbox{\rm O}(N,\CM)\;
\right\}
\;\cong\;
\Gg^{\mbox{\rm\tiny D}}
\;.
$$
Thus the results of Section~\ref{sec-LyapD} apply. In particular, the Lyapunov spectrum of $H_K$ is always twice degenerate (because the spectrum in $\mbox{\rm O}(N,\CM)$ is symmetric), and there is always a twice degenerate vanishing Lyapunov exponent if $N$ is odd.

\begin{comment}

The $H$ Class DIII has a chiral symmetry so that one is tempted to use again the model $H_I$ with SLS $J^*H_IJ=-H_I$. However, this will not work because this SLS is even WRONG BECAUSE NEED NOT COMMUTE, TRY AGAIN!, and the one in Class DIII has to be odd. Therefore let us work with the model Hamiltonian $H_J$ and rather implement the (odd) SLS by $I^*H_JI=-H_J$. Then the even PHS is $\overline{H}_J=-H_J$ which then implies the odd TRS $I^*\overline{H_J}I=H_J$. Now one checks that the transfer matrices satisfy $\overline{T_n}=T_n$ and also $I^*\overline{T_n}I=T_n$, which combined also shows $I^*T_n I=T_n$. Hence
%
$$
\Tt_J^{\mbox{\rm\tiny DIII}}\;=\;
\left\{
T\in\mbox{\rm GL}(2N,\CM)\,\left|\,
T^*J \,T\,=\,J\;,\; \overline{T}\,=\,T\,,\;\;I^*\,T\,I\,=\,T
\right.
\right\}
\;.
$$
%
Taking Cayley transform, one then finds
%
$$
C\,\Tt_J^{\mbox{\rm\tiny DIII}}\,C^*\;=\;
\left\{
\left.
\left(\begin{array}{cc} A & 0 \\ 0 & (A^{-1})^* \end{array}\right)
\,\right|\,
A\in \mbox{\rm O}(N,\CM)\;
\right\}
\;\cong\;
\Gg^{\mbox{\rm\tiny D}}
\;.
$$
%
Thus the results of Section~\ref{sec-LyapD} apply. In particular, the Lyapunov spectrum of $H_I$ is always twice degenerate (because that in $\mbox{\rm O}(N,\CM)$ is symmetric) and there is always a twice degenerate vanishing Lyapunov exponent if $N$ is odd.

\end{comment}

%%%%%%%%%%%%%%%%%%%%%%%%%%%%%%%%%%%%%%%%%%%%%%%%%%%%%%%%%%%%%%%%%%%%
\subsection{Odd PHS and even TRS ($H$ Class CI)}

Again the odd PHS enforces the use of a supplementary grading, so the Hilbert space is $L^2(\RM,\CM^{2N})\otimes \CM^2$. Let us choose the model operator $H_K$ and implement the even TRS by $I\otimes I\, \overline{H_K}\, I\otimes I=H_K$ and the odd PHS $K\otimes I^* \,\overline{H_K}\, K\otimes I=-H_K$. Then also the SLS $J\otimes \one \,H_K \,J\otimes\one= -H_K$ holds. Next one verifies that the transfer matrices are in the following group:
$$
\Tt_K^{\mbox{\rm\tiny CI}}\,=\,
\left\{
T\in\mbox{\rm GL}(4N,\CM)\,\left|\,
T^*K\otimes\one\, T=K\otimes\one\,,\; K\otimes I^*\,\overline{T}\,K\otimes I=T\,,\;J\otimes \one\,T\,J\otimes \one=T
\right.
\right\}
\,.
$$
After some algebra,  the latter relations show
$$
\Tt_K^{\mbox{\rm\tiny CI}}\;=\;
\left\{
\left.
\left(\begin{array}{cc} A & 0 \\ 0 & (A^{-1})^* \end{array}\right)
\,\right|\,
A\in \mbox{\rm SP}(2N,\CM)\;
\right\}
\;\cong\;
\Gg^{\mbox{\rm\tiny C}}
\;.
$$
Hence the results of Section~\ref{sec-LyapC} apply. In particular, the Lyapunov spectrum of $H_K$ is always twice degenerate (because that in $\mbox{\rm SP}(2N,\CM)$ is symmetric).

%%%%%%%%%%%%%%%%%%%%%%%%%%%%%%%%%%%%%%%%%%%%%%%%
\section{Symmetry enforced delocalization}
\label{sec-deloc}

The paper \cite{SS} studies the model Hamiltonian $H_I$ of \eqref{eq-Ham} in the $H$ Class AII and, in particular, the case where the channel number $N$ is odd. As explained in Section~\ref{sec-AII} there are at least two vanishing Lyapunov exponents in this case. Furthermore, if a certain coupling hypothesis formulated in \cite{SS} holds, there are exactly two vanishing Lyapunov exponents, just as in the case with RPP.  By Kotani theory for Dirac operators \cite{Sun,SS}  the two vanishing Lyapunov exponents imply that the spectrum of $H$ has almost surely an absolutely continuous component of multiplicity $2$. Furthermore, it was possible to show in \cite{SS} that  almost surely there is no singular spectrum, provided the distribution of the $\Vv_n$ is absolutely continuous. All these results directly transpose to $H$ Class DIII in the case where $N$ is odd. No details of proof are provided here because there is really no essential difference except for the verification that there are at most $2$ non-vanishing Lyapunov exponents in a situation with sufficient coupling. 

\vspace{.2cm}

Finally let us discuss systems with a different number of left and right movers which are relevant for the modeling of the edge modes in various quantum Hall systems:
$$
H_G\;=\;
G\,\imath\,\partial
\,+\,
\Vv\;,
$$

\noindent acting on $L^2(\RM,\CM^{N+M})$ where $N\leq M$ are integers. If $H$ satisfies a TRS or a SLS, then $N=M$ because otherwise the kinetic part is not isospectral to its negative. But in the remaining $H$ Classes A,C and D it is permitted to have $M>N$, see Table~1. It follows that the transfer matrices are in the group $\mbox{\rm U}(N,M)$. For the $H$ Classes C and D they are actually in the subgroups $\mbox{\rm O}(N,M)$ and $\mbox{\rm SP}(\frac{N}{2},\frac{M}{2})$, where $N$ and $M$ have to be even in the latter case. Now the spectral theory of a self-adjoint $T\in\mbox{\rm U}(N,M)$ shows that $T$ has at least $M-N$ eigenvalues $1$, see Section~\ref{sec-GAIII}.  Thus $H_G$ must have $M-N$ vanishing Lyapunov exponents. Under a coupling hypothesis on the random potential similar as in \cite{SS},  there are no other vanishing Lyapunov exponents. Again it follows from Kotani theory alluded to above that these Hamiltonians have almost surely an absolutely continuous spectrum of multiplicity $M-N$. Again one can show that this absolutely continuous spectrum is almost surely pure if the distribution of the $\Vv_n$ is absolutely continuous.

\vspace{.5cm}

\noindent {\bf Acknowledgements:} We are thankful for financial support by the DFG. 
The basic idea to write this paper resulted from discussions at an SFB/TR 12 meeting in 
Gda\'nsk in 2009. We all thank M.~Zirnbauer for the invitation. H.~S.-B. also thanks the 
Mittag-Leffler Institute for a quiet office space which finally allowed to finish this work. This
work was supported, in part, by the NSF under grant
DMR-0706140 (A.W.W.L.).

\appendix

%%%%%%%%%%%%%%%%%%%%%%%%%%%%%%%%%%%%%%%%%%
\section{Moments for the Haar measures over compact groups}

Section \ref{App-results}  of this appendix collects formulas for certain averages over the compact classical groups that are used in the main part of this paper. These formulas can be derived rather easily from results scattered over the literature, and those involving averages over the unitary group were already listed in \cite{RS}. Nevertheless, it is emphasized in the remaining sections of the appendix that all these formulas follow from a useful calculus involving the so-called Weingarten function \cite{Wei, CS, CSt}, which may be viewed as a systematic approach to invariant theoretic arguments in the physics literature on averages, as in Section VIII of \cite{AZ}. To illustrate how invariant theory is applied, Section \ref{App-sympl} deals explicitly with the fourth moments over the symplectic group (averages of a product of four matrix entries) and provides the corresponding value of the Weingarten function. This allows in Section~\ref{Wg-appl} to deduce Lemma~\ref{lem-SPmoments}(ii), which is actually the most complicated of all formulas in Section~\ref{App-results}. All others follow similarly from prior results on the orthogonal group \cite{CS}. Higher moments can also be calculated explicitly, provided that the matrix size is sufficiently large compared to the order of the moment (that is, the number of matrix entries that are being considered). Section \ref{Wg-general} presents the general version of the Weingarten integration formula for the unitary, orthogonal, and symplectic groups. These results are not really used in this paper, but they are definitely of independent interest and are provided for further reference.

%%%%%%%%%%%%%%%%%%%%%%%%%%%%%%%%%%%%%%%%%%
\subsection{Collection of formulas used in Section~\ref{sec-Lyapspec}}
\label{App-results}

%%%%%%%%%%%%%%%%%%%%%%%%%%%%%%%%%%%%%%%%%%%%
\begin{lemma}
\label{lem-Umoments} Let $A,B,C,D\in\mbox{\rm Mat}(N\times N,\CM)$. The following holds for averages over $\mbox{\rm U}(N)$:
\begin{eqnarray*}
\mbox{\rm (i) }\;\;\;\;\;\;\;\;\;\;\;\;\;\;\;\;
\langle\,\Tr(U^*AUB)\,\rangle
& = &
\frac{1}{N}\,\Tr(A)\Tr(B)\;,
\\
\mbox{\rm (ii) }\;\;\;\;\;\;\;\;\;\;\;\;\;\;\;\;\;
\langle\,\Tr(\overline{U}AUB)\,\rangle
& = &
\frac{1}{N}\,\Tr(AB^t)\;,
\qquad
\langle\,\Tr(U^tAUB)\,\rangle
\; = \;0\;,
\\
\mbox{\rm (iii) }\;\,\;
\langle\,\Tr(U^*AUBU^*CUD)\,\rangle
& = &
\frac{1}{N^2-1}\Big[\Tr(A)\Tr(C)\Tr(BD)+\Tr(AC)\Tr(B)\Tr(D)\Big]
\\
& & \!\!\!\!\!\!\!\!-\;
\frac{1}{N(N^2-1)}\Big[\Tr(AC)\Tr(BD)+\Tr(A)\Tr(B)\Tr(C)\Tr(D)\Big]
\;,
\\
\mbox{\rm (iv) }\;\;
\langle\,\Tr(U^*AUBU^tC\overline{U}D)\,\rangle
& = &
\frac{1}{N^2-1}\Big[\Tr(A)\Tr(C)\Tr(BD)+\Tr(AC^t)\Tr(BD^t)\Big]
\\
& & -\;
\frac{1}{N(N^2-1)}\Big[\Tr(AC^t)\Tr(BD)+\Tr(A)\Tr(C)\Tr(BD^t)\Big]
\;,
\\
\mbox{\rm (v) }\;\;
\langle\,\Tr(U^*A\overline{U}BU^tCUD)\,\rangle
& = &
\frac{1}{N^2-1}\Big[\Tr(AC^t)\Tr(BD^t)+\Tr(AC)\Tr(B)\Tr(D)\Big]
\\
& & -\;
\frac{1}{N(N^2-1)}\Big[\Tr(AC)\Tr(BD^t)+\Tr(AC^t)\Tr(B)\Tr(D)\Big]
\;.
\end{eqnarray*}
All formulas remain valid if all $U$'s are replaced by their complex conjugates, {\it e.g.}
$$
\langle\,\Tr(U^tA\overline{U}BU^tC\overline{U}D)\,\rangle
\;=\;
\langle\,\Tr(U^*{A}U{B}U^*{C}{U}{D})\,\rangle
\;.
$$
\end{lemma}
%%%%%%%%%%%%%%%%%%%%%%%%%%%%%%%%%%%%%%%%%%%%

%\vspace{.2cm}

%%%%%%%%%%%%%%%%%%%%%%%%%%%%%%%%%%%%%%%%%%%%
\begin{lemma}
\label{lem-Omoments} Let $A,B,C,D\in\mbox{\rm Mat}(N\times N,\CM)$. The following holds for averages over $\mbox{\rm O}(N)$:
\vspace{.1cm}

\noindent {\rm (i)}
$$
\langle\,\Tr(O^tAOB)\,\rangle
\; = \;
%\mbox{\rm Wg}([1])
\frac{1}{N}\,\Tr(A)\Tr(B)
\;,
\qquad
\langle\,\Tr(OAOB)\,\rangle
\; = \;
%\mbox{\rm Wg}([1])
\frac{1}{N}\,\Tr(A B^t)
\;.
$$
{\rm (ii) }
\begin{eqnarray*}
& & 
\hspace{-1cm}
\langle\,\Tr(O^tAOBO^tCOD)\,\rangle
\\
& = &
\frac{N+1}{N(N-1)(N+2)}
\,\Big[\Tr(A)\Tr(C)\Tr(BD)+\Tr(AC^t)\Tr(BD^t)+\Tr(AC)\Tr(B)\Tr(D)\Big]
\\
& & 
-\frac{1}{N(N-1)(N+2)}
\,\Big[\Tr(A)\Tr(C)\Tr(BD^t)+\Tr(A)\Tr(C)\Tr(B)\Tr(D)+ \Tr(AC^t)\Tr(BD)
\\
& & \;\;\;\;\;\;\;\;\;\;\;\;\;\;\;\;\;\;
+\,\Tr(AC^t)\Tr(B)\Tr(D)
+\Tr(AC)\Tr(BD)+\Tr(AC)\Tr(BD^t)
\Big]
\;.
\end{eqnarray*}
{\rm (iii) }
\begin{eqnarray*}
& &
\hspace{-1cm} 
\langle\,\Tr(O^tAO^tBOCOD)\,\rangle
\\
& = &
\frac{N+1}{N(N-1)(N+2)}\,\Big[\Tr(A^tBC^tD)+\Tr(DCBA)+\Tr(AC)\Tr(B)\Tr(D)\Big]
\\
& & 
-\,\frac{1}{N(N-1)(N+2)}
\,\Big[\Tr(A^tBC^tD^t)+\Tr(A^tBC^t)\Tr(D)+\Tr(CBAD^t)
\\
& & \;\;\;\;\;\;\;\;\;\;\;\;\;\;\;\;\;\;
+\,\Tr(CBA)\Tr(D)
+\Tr(CAD^t)\Tr(B)+\Tr(CAD^t)\Tr(B)
\Big]
\;.
\end{eqnarray*}

\end{lemma}
%%%%%%%%%%%%%%%%%%%%%%%%%%%%%%%%%%%%%%%%%%%%

%%%%%%%%%%%%%%%%%%%%%%%%%%%%%%%%%%%%%%
\begin{lemma}
\label{lem-SPmoments} Let $A,B,C,D\in\mbox{\rm Mat}(2N\times 2N,\CM)$. The following holds for averages over $\mbox{\rm SP}(2N)$:
\noindent {\rm (i)} One has
\begin{equation}
\label{eq-Sym2nd}
\langle\Tr(U^*AUB)\rangle
\;=\;
\frac{1}{2N}
\,\Tr(A)\,\Tr(B)
\;,
\qquad
\langle\Tr(U^tAUB)\rangle
\;=\;
\frac{1}{2N}
\,\Tr(IA)\,\Tr(I^*B)
\;,
\end{equation}

and
$$
\langle\Tr(UAUB)\rangle
\;=\;
\langle\Tr(U^*AU^*B)\rangle
\;=\;
\frac{1}{2N}
\,\Tr(AIB^tI)\;,
\qquad
\langle\Tr(\overline{U}AUB)\rangle
\;=\;
\frac{1}{2N}
\,\Tr(AB^t)
\;.
$$

\noindent {\rm (ii)}
\begin{eqnarray*}
& & 
\hspace{-.9cm} \langle\,\Tr(U^*AUBU^*CUD)\,\rangle
\\
& = &
\frac{1}{4N(N-1)(2N+1)}\;
\Bigl[
(2N-1)\Tr(A)\Tr(C)\Tr(BD)
+\Tr(A)\Tr(C)\Tr(I^*B^tID)
\\
& &
-\Tr(A)\Tr(C)\Tr(B)\Tr(D)
-\Tr(I^*A^tIC)\Tr(BD)
+\Tr(I^*A^tIC)\Tr(B)\Tr(D)
\\
& &
-\Tr(AC)\Tr(BD)
-(2N-1)\Tr(I^*A^tIC)\Tr(I^*B^tID)
\\
& &
-\Tr(AC)\Tr(I^*B^tID)
+(2N-1)\Tr(AC)\Tr(B)\Tr(D)
\Bigr]
\;.
\end{eqnarray*}
\end{lemma}
%%%%%%%%%%%%%%%%%%%%%%%%%%%%%%%%%%%%%%

Let us close this section by stating two results that follow from the above and are also used in Section~\ref{sec-reviewclassical}. Recall from Section~\ref{sec-GCI} the special form of $\Uu^{\mbox{\rm\tiny CI}}=\Uu^{\mbox{\rm\tiny DIII}}\subset\mbox{\rm U}(2N)$, which is is isomorphic to $\mbox{\rm U}(N)$. The following lemma presents no general formula for second and fourth moments of this group, but only those moments which are needed in Section~\ref{sec-LyapDIII}. The proof of the lemma is a calculation based on the identities in Lemma~\ref{lem-Umoments}.

%%%%%%%%%%%%%%%%%%%%%%%%%%%%%%%%%%%%%%
\begin{lemma}
\label{lem-UCImoments} Let $A,B\in\mbox{\rm Mat}(2N\times 2N,\CM)$.  Then the averages over $\Uu^{\mbox{\rm\tiny CI}}=\Uu^{\mbox{\rm\tiny DIII}}$  lead to the following:
\vspace{.1cm}

\noindent {\rm (i)} With $\Pi_+=\binom{\one \;0}{0\;0}$ and $\Pi_-=\binom{0 \;0}{0\;\one}$
$$
\langle\,\Tr(U^*AUB)\,\rangle
\;=\;
\frac{1}{N}
\;\Tr
\left(
\Pi_+A\right)
\Tr
\left(\Pi_+B\right)
\;+\;
\frac{1}{N}
\;\Tr
\left(\Pi_-A\right)
\Tr
\left(\Pi_-B\right)
\;.
$$
\noindent {\rm (ii)} 
Let $A$ and $B$ be of the form
$$
A
\;=\;
\begin{pmatrix}
0 & a \\ a^* & 0
\end{pmatrix}
\;,
\qquad
B
\;=\;
\begin{pmatrix}
b & 0 \\ 0 & c
\end{pmatrix}
\;,
$$

where all entries are of size $N\times N$. Then
$$
\langle\,\Tr(U^*AUBU^*AUB)\,\rangle
 = 
\frac{2\Tr(a^*a)(N\Tr(b)\Tr(c)-\Tr(bc^t))
+
2\Tr(\overline{a}a)(N\Tr(bc^t)-\Tr(b)\Tr(c))}{N(N^2-1)}.
$$
\noindent {\rm (iii)} If, moreover, $a^t=-a$, then
$$
\langle\,\Tr(U^*AUBU^*AUB)\,\rangle
\; = \;
\frac{\Tr(A^2)(\Tr(b)\Tr(c)-\Tr(bc^t))}{N(N-1)}\;.
$$
\end{lemma}
%%%%%%%%%%%%%%%%%%%%%%%%%%%%%%%%%%%%%%

Recall from Section~\ref{sec-GCI} the special form of $\Uu^{\mbox{\rm\tiny CII}}=\mbox{\rm SP}(2N)\times \mbox{\rm SP}(2N)$. The following lemma presents no general formula for second and fourth moments of this group, but only those moments which are needed in Section~\ref{sec-LyapCII}. The proof of the lemma is a somewhat tedious calculation based on the identities in Lemma~\ref{lem-SPmoments}(i).

%%%%%%%%%%%%%%%%%%%%%%%%%%%%%%%%%%%%%%
\begin{lemma}
\label{lem-UCIImoments} Let $A,B\in\mbox{\rm Mat}(4N\times 4N,\CM)$.  Then the averages over $\Uu^{\mbox{\rm\tiny CI}}$  lead to the following:
\vspace{.1cm}

\noindent {\rm (i)} With $\Pi_+=\binom{\one \;0}{0\;0}$ and $\Pi_-=\binom{0 \;0}{0\;\one}$ as above,
$$
\langle\,\Tr(U^*AUB)\,\rangle
\;=\;
\frac{1}{2N}
\;\Tr
\left(
\Pi_+A\right)
\Tr
\left(\Pi_+B\right)
\;+\;
\frac{1}{2N}
\;\Tr
\left(\Pi_-A\right)
\Tr
\left(\Pi_-B\right)
\;.
$$
\noindent {\rm (ii)} 
Let $A$ and $B$ be of the form
$$
A
\;=\;
\begin{pmatrix}
0 & a \\ a^* & 0
\end{pmatrix}
\;,
\qquad
B
\;=\;
\begin{pmatrix}
e & f \\ f & e
\end{pmatrix}
\;,
$$

where all entries are of size $2N\times 2N$ and furthermore $I\overline{a}I=a$ and $B^*=B^t=B$. Then
$$
\langle\,\Tr(U^*AUBU^*AUB)\,\rangle
\; = \;
\frac{\Tr(A^2)\bigl(\Tr(e)^2+\Tr(I^*fIf)\bigr)}{4\,N^2}\;.
$$
\end{lemma}
%%%%%%%%%%%%%%%%%%%%%%%%%%%%%%%%%%%%%%

%%%%%%%%%%%%%%%%%%%%%%%%%%%%%%%%%%%%%%%%%%%%
\subsection{The symplectic Weingarten function for fourth moments}
\label{App-sympl}

As already pointed out, the formulas of Section~\ref{App-results} can be deduced from \cite{Co,CS} except for averages involving the symplectic group. In the symplectic case a sign factor is missing in \cite{CS}. This was corrected in \cite{CSt}, but no explicit value of the Weingarten function was given. This section treats the fourth moment needed for the proof of Lemma~\ref{lem-SPmoments}(ii) in detail. This also indicates how in principle the general results stated in Section~\ref{Wg-general} can be proved.

\vspace{.2cm}

Let $e_j$, $j=1,\ldots, 2N$, denote the standard basis of the complex vector space $V=\CM^{2N}$. For any linear map $T$ on $V$, the matrix entries are denoted by $T_{i,j}=e_i^*Te_j$. Associated to $I$ defined in \eqref{eq-symmdef}, let us introduce on $V$ the skew symmetric bilinear form $a(v,w)=v^t Iw$ where $v^t=\overline{v}^*$ denotes the transpose of $v\in V$. Then for $i, j = 1, \ldots, N$,
\begin{eqnarray*}
T_{i + N, j}\, =\, - \, a(e_i, T e_j )\,, 
\;\; & & 
T_{i+N, j+N} \,= \, - \,a(e_i, T e_{j+N} )\,,
\\
\;\;
T_{i, j}\, =\, a(e_{i+N}, T e_{j} )\,,
\;\; & &
T_{i, j+N} \,=\, 
a(e_{i+N}, T e_{j+N})
\,.
\end{eqnarray*}
Thus, for $i_1, j_1, \ldots, i_4, j_4 \in \{ 1, \ldots, N\}$ and $\alpha_1, \beta_1, \ldots, \alpha_4, \beta_4 \in \{0, 1\}$, 
\begin{align}
\label{eq-invariantstart} 
&\int_{\on{SP}(2N)} \!\!\!dU\;
U_{i_1 + \alpha_1 N, j_1 + \beta_1 N}\, U_{i_2 + \alpha_2 N, j_2 + \beta_2 N}\, U_{i_3 + \alpha_3 N, j_3 + \beta_3 N}\,U_{i_4 + \alpha_4 N, j_4 + \beta_4 N}
\\
&= \,(-1)^{\alpha_1 + \alpha_2 + \alpha_3 + \alpha_4}\, 
a\left( e_{i_1 + (1 - \alpha_1) N} \otimes \ldots \otimes 
e_{i_4 + (1 - \alpha_4) N},\int_{\on{SP}(2N)} \!\!\!dU\; U e_{j_1 + \beta_1 N} \otimes 
\ldots \otimes U e_{j_4 + \beta_4 N}\right), 
\nonumber
\end{align}
where $a$ is extended to $V^{\otimes 4}$ via
$$ 
a(v_1 \otimes v_2  \otimes v_3 \otimes v_4, 
w_1 \otimes w_2  \otimes w_3 \otimes w_4) 
\;=\; 
a(v_1, w_1) \,a(v_2, w_2)\,a(v_3, w_3)\, a(v_4, w_4)\,.
$$ 
Note that this yields a symmetric bilinear form on $V^{\otimes 4}$. By the definition of Haar measure, the integral
\begin{equation}
\label{proj-invar}
\int_{\on{SP}(2N)} \!\!\!dU\; U e_{j_1 + \beta_1 N} \otimes 
\ldots \otimes U e_{j_4 + \beta_4 N}
\end{equation}
is a $\on{SP}(2N)$-invariant in $V^{\otimes 4}$, specifically, it is fixed under the action of $U \in \on{SP}(2N)$ on $V^{\otimes 4}$ which on decomposable tensors is given via $U(v_1 \otimes v_2 \otimes v_3 \otimes v_4) = (U v_1 \otimes U v_2 \otimes U v_3 \otimes U v_4).$ If $N \ge 2$, then by the symplectic case of Weyl's First Fundamental Theorem for tensor invariants \cite[Thm 5.3.3]{GW} and  the results of Section 3.4 in \cite{St}, a basis of the (three-dimensional) subspace of invariant tensors in $V^{\otimes 4}$ can be described as follows. Write $\frp_1 = \{ \{1, 2\}, \{3, 4\}\}$, $\frp_2 = \{ \{1,3\}, \{2, 4\}\}$ and $\frp_3 = \{ \{1, 4\}, \{2, 3\}\}$. These are the pair partitions of the set $\{ 1, 2, 3, 4\}$. To control the sign factors that arise from the skew symmetry of the form $a$ on $V$, let us keep track of the natural order on the individual blocks by using the following notation.  For  $\frm \in \{\frp_1, \frp_2, \frp_3\}$, let us write $\frm = \{ (m_1(\frm), n_1(\frm)), (m_2(\frm), n_2(\frm)) \}$, where $\{m_1(\frm), n_1(\frm)\}, \{m_2(\frm), n_2(\frm)\}$ are the blocks of $\frm$, and $m_1(\frm) < n_1(\frm),  \;m_2(\frm) < n_2(\frm)$.  To $\frp_1$ is now associated the tensor
$$ 
\theta_1 
\;=\;
\theta_{\frp_1} 
\;=\; 
\sum_{\eta_1, \eta_2 = 1}^{N}\,
\sum_{\epsilon_1, \epsilon_2 = 0}^1 \, 
e_{\eta_1 + \epsilon_1 N}
\otimes e_{\eta_1 + (1 - \epsilon_1)N} (-1)^{\epsilon_1} \otimes
e_{\eta_2 + \epsilon_2 N}
\otimes e_{\eta_2 + (1 - \epsilon_2)N} (-1)^{\epsilon_2}
\;.
$$
Here, the maps $r \mapsto \eta_r$ and $r \mapsto \epsilon_r$ are constant on the block $\{ m_1(\frp_1), n_1(\frp_1)\}$ as well as on  $\{ m_2(\frp_1), n_2(\frp_1)\}$, and to the $m$ component of a block corresponds a vector of the form $e_{\eta_r + \epsilon_r N}$,  while to the $n$ component corresponds a vector of the form $(-1)^{\epsilon_r} e_{\eta_r + (1 - \epsilon_r)N}$. Analogously, 
\begin{align*}
 \theta_2 &\;= \;\theta_{\frp_2} \;= \;
 \sum_{\eta_1, \eta_2 = 1}^{N}\,
\sum_{\epsilon_1, \epsilon_2 = 0}^1  \,
e_{\eta_1 + \epsilon_1 N}
\otimes
e_{\eta_2 + \epsilon_2 N}
\otimes e_{\eta_1 + (1 - \epsilon_1)N} (-1)^{\epsilon_1} 
\otimes e_{\eta_2 + (1 - \epsilon_2)N} (-1)^{\epsilon_2}\;,
\\ 
\theta_3 &\;=\; \theta_{\frp_3}\; =\;
\sum_{\eta_1, \eta_2 = 1}^{N}\,
\sum_{\epsilon_1, \epsilon_2 = 0}^1\,  
e_{\eta_1 + \epsilon_1 N}
 \otimes
e_{\eta_2 + \epsilon_2 N}
\otimes e_{\eta_2 + (1 - \epsilon_2)N} (-1)^{\epsilon_2}
\otimes e_{\eta_1 + (1 - \epsilon_1)N} (-1)^{\epsilon_1}
\;.
\end{align*}
Then $\{ \theta_1, \theta_2, \theta_3\}$ is a basis of the space of symplectic invariants in $V^{\otimes 4}$, and it is a straightforward (tedious) computation to verify that the Gram matrix of the form $a$ with respect to this basis is given by
$$ 
\left( \begin{array}{lll} 4 N^2 & 2N & - 2N\\
2N & 4 N^2 & 2N\\ - 2N & 2N & 4 N^2\end{array} \right)
\;.
$$
As long as $N \ge 2$, this matrix is invertible with inverse 
\begin{equation}
\label{Wg-4}
\frac{1}{4N (N-1) (2N+1)} \left( \begin{array}{lll}
2N - 1 & -1 & 1 \\ -1 & 2N - 1 & -1 \\ 1 & -1 & 2N - 1
\end{array} \right)
\;.
\end{equation}
By definition, the $(i, j)$-entry of this inverse is the value $\on{Wg}_{\on{SP}(2N)}(\frp_i, \frp_j)$ of the symplectic Weingarten function. In Section~\ref{Wg-general}, the general Weingarten function $\on{Wg}_{\on{SP}(2N)}$ is defined by the matrix entries of the inverse of the Gram matrix of $a$ w.r.t. a basis of the invariants, which in turn is in bijection with the pair partitions (of $2k$ points if $2k$th moments are considered). In order to evaluate \eqref{eq-invariantstart}, one now expresses the invariant \eqref{proj-invar} in terms of the basis $\{ \theta_1, \theta_2, \theta_3\}$. Noting that $a( e_{i_1 + \alpha_1N} \otimes \ldots \otimes e_{i_4 + \alpha_4 N}, \theta_{\frp_j})$ vanishes unless $i_{m_r(\frp_j)} = i_{n_r(\frp_j)}$ and  $\alpha_{m_r(\frp_j)} = 1 - \alpha_{n_r(\frp_j)}$ for $r = 1, 2$, one obtains
%
%\label{int-sympl}
\begin{align}
\int_{\on{SP}(2N)} \!\!\!dU\;
&
U_{i_1 + \alpha_1 N, j_1 + \beta_1 N}\,U_{i_2 + \alpha_2 N, j_2 + \beta_2 N}\,U_{i_3 + \alpha_3 N, j_3 + \beta_3 N}\, U_{i_4 + \alpha_4 N, j_4 + \beta_4 N}
\label{eq-Sp4th}
\\
 =\; & 
 \sum_{\frm, \frn \in \{ \frp_1, \frp_2, \frp_3\}} 
\on{Wg}_{\on{SP}(2N)}(\frm, \frn)\
(-1)^{\alpha_{m_1(\frm)} + \alpha_{m_2(\frm)} + \beta_{m_1(\frn)} +
\beta_{m_2(\frn)}}
\nonumber
\\ & \hspace{4em}
\delta({i_{m_1(\frm)}, i_{n_1(\frm)}})\ \delta({\alpha_{m_1(\frm)}, 1 - \alpha_{n_1(\frm)}})\  
\delta({i_{m_2(\frm)}, i_{n_2(\frm)}})\ \delta({\alpha_{m_2(\frm)}, 1 - \alpha_{n_2(\frm)}})
\nonumber
\\ 
& \hspace{4em}
\delta({j_{m_1(\frn)}, j_{n_1(\frn)}})\ \delta({\beta_{m_1(\frn)}, 1 - \beta_{n_1(\frn)}})\ 
\delta({j_{m_2(\frn)}, j_{n_2(\frn)}})\ \delta({\beta_{m_2(\frn)}, 1 - \beta_{n_2(\frn)}})
\;. 
\nonumber
\end{align}

%%%%%%%%%%%%%%%%%%%%%%%%%%%%%%%%%%%%%%%%%%%%%%
\subsection{Proof of Lemma~\ref{lem-SPmoments}(ii)}
\label{Wg-appl}

Let us begin by writing out the l.h.s. explicitly:
\begin{align*}
&\left\langle \on{Tr}( I U^t I A U B I U^t I C U D) \right\rangle\\
  &\;\;\;=\; \sum_{i_1, i_2, \ldots, i_8 = 1}^{N} \sum_{\alpha_1, \alpha_2, \ldots, \alpha_8 = 0}^1 I_{i_1 + \alpha_1 N, i_1 + (1- \alpha_1)N}\,
 I_{i_2 + (1 - \alpha_2)N, i_2 + \alpha_2 N}\, A_{i_2 + \alpha_2 N,
 i_3 + \alpha_3 N}\,B_{i_4 + \alpha_4 N, i_5 + \alpha_5 N}\\
 &
 \hspace{2cm} 
 I_{i_5 + \alpha_5 N, i_5 + (1 - \alpha_5)N}\, 
 I_{i_6 + (1-\alpha_6)N, i_6 + \alpha_6 N}\,C_{i_6 + \alpha_6 N,
 i_7 + \alpha_7 N}\,D_{i_8 + \alpha_8 N, i_1 + \alpha_1 N} 
 \\
 &
 \hspace{2cm} 
\left\langle U_{i_2 + (1 - \alpha_2)N, i_1 + (1 - \alpha_1)N}\,U_{i_3 + \alpha_3 N, i_4 + \alpha_4 N}\, U_{i_6 + (1 - \alpha_6) N, i_5 + (1 - \alpha_5)N}\,U_{i_7 + \alpha_7 N, i_8 + \alpha_8 N}\right\rangle.
\end{align*}
The next step is to expand this average $\langle\;.\;\rangle=\int_{\on{SP}(2N)} \!dU$ according to \eqref{eq-Sp4th}. Then one interchanges the sum over pairs of pair partitions with the sums over the $i$ and $\alpha$ indices and then determines the contribution of each pair $(\frp_i, \frp_j)$. Let us illustrate this procedure for the pair $(\frp_1, \frp_2)$. One reads off from \eqref{eq-Sp4th}  that for a choice of $i$ and $\alpha$ indices to give a non-vanishing contribution, one must have that $i_3 = i_2,\; \alpha_3 = \alpha_2, \;i_7 = i_6, \;\alpha_7 = \alpha_6,\; i_5 = i_1,\; \alpha_5 = 1 - \alpha_1, \;i_8 = i_4,\; \alpha_8 = 1-\alpha_4$. Furthermore, the sign factor corresponding to each choice of $i$ and $\alpha$ indices is 
$$ 
(-1)^{(1-\alpha_2) + (1-\alpha_6) + (1 - \alpha_1) + \alpha_4}
\;=\; 
(-1)^{\alpha_1 + \alpha_2 + \alpha_4 + (1-\alpha_6)}
\;.
$$ 
Consequently, the coefficient of $\on{Wg}_{\on{SP}(2N)}(\frp_1, \frp_2)$ in the sum over pairs of pair partitions is
\begin{align*}
\sum_{i, \alpha} (-1)^{\alpha_1 + \alpha_2 + \alpha_4 + (1-\alpha_6)}\
&I_{i_1 + \alpha_1 N, i_1 + (1 - \alpha_1)N}\, 
I_{i_2 + (1 - \alpha_2)N, i_2 + \alpha_2 N}\,A_{i_2 + \alpha_2 N, i_2 + \alpha_2 N}\,B_{i_4 + \alpha_4N, i_1 + (1-\alpha_1)N}
\\
&I_{i_1 + (1 - \alpha_1)N, i_1 + \alpha_1 N}\,
I_{i_6 + (1 - \alpha_6)N, i_6 + \alpha_6 N}\, 
C_{i_6 + \alpha_6 N, i_6 + \alpha_6 N}\,
D_{i_4 + (1 - \alpha_4)N, i_1 + \alpha_1 N}
\;.
\end{align*}
Observing that $I_{i + \alpha N, i + (1-\alpha)N} = (-1)^{1-\alpha}$ and $I_{i + (1 - \alpha)N, i+ \alpha N} = (-1)^{\alpha}$, one can group the factors as follows:
\begin{align*}
&A_{i_2 + \alpha_2 N, i_2 + \alpha_2 N}\;,\\
&C_{i_6 + \alpha_6 N, i_6 + \alpha_6 N}\;,\\
&B_{i_4 + \alpha_4N, i_1 + (1-\alpha_1)N}\, I_{i_1 + (1 - \alpha_1)N, i_1 + \alpha_1 N}\,
D_{i_4 + (1 - \alpha_4)N, i_1 + \alpha_1 N}\
(-1)^{\alpha_4}
\\
&\;\;\;=\;
B_{i_4 + \alpha_4N, i_1 + (1-\alpha_1)N}\,
I_{i_1 + (1 - \alpha_1)N, i_1 + \alpha_1 N}\,
D_{i_4 + (1 - \alpha_4)N, i_1 + \alpha_1 N}\,
I_{i_4 + (1 - \alpha_4)N, i_4 + \alpha_4 N}\;,
\\
&
I_{i_1 + \alpha_1 N, i_1 + (1 - \alpha_1)N} (-1)^{\alpha_1} \;=\; -1\,,
\\
& 
I_{i_2 + (1 - \alpha_2)N, i_2 + \alpha_2 N} (-1)^{\alpha_2}\; =\; 1\,,
\\
& 
I_{i_6 + (1 - \alpha_6)N, i_6 + \alpha_6 N} (-1)^{1 - \alpha_6} 
\;=\; 
-\,1\;.
\end{align*}
Hence, $\on{Wg}_{\on{SP}(2N)}(\frp_1, \frp_2)$ comes with the coefficient $\on{Tr}(A) \on{Tr}(C) \on{Tr}(BID^tI)$. Going in this manner through all nine cases yields
\begin{align*}
& \left\langle \on{Tr} ( U^{-1} A U B U^{-1} C U D) \right\rangle 
\; = \;\left\langle \on{Tr} ( I U^t I A U B I U^t I C U D) \right\rangle\\
& \;\;\;= \; \on{Wg}(\frp_1, \frp_1) \on{Tr}(A) \on{Tr}(C) \on{Tr}(BD)
\,+\, \on{Wg}(\frp_1, \frp_2) \tr(A) \tr(C) \tr(B I D^t I )
\\
& \;\;\;\;\;\;\;\;\;
- \on{Wg}(\frp_1, \frp_3) \tr(A) \tr(B) \tr(C) \tr(D)
\,-\, \on{Wg}(\frp_2, \frp_1) \tr(AIC^tI) \tr(BD)\\
&
\;\;\;\;\;\;\;\;\;- \,\on{Wg}(\frp_2, \frp_2) \tr(A I C^t I) \tr(B I D^t I)
\,+\, \on{Wg}(\frp_2, \frp_3) \tr(A^t I C I ) \tr(B) \tr(D) \\
&
\;\;\;\;\;\;\;\;\; - \,\on{Wg}(\frp_3, \frp_1) \tr(AC) \tr(BD)
\,-\, \on{Wg}(\frp_3, \frp_2) \tr(AC) \tr(BID^tI) 
\\
&
\;\;\;\;\;\;\;\;\;
+ \,\on{Wg}(\frp_3, \frp_3) \tr(AC) \tr(B) \tr(D)\;,
\end{align*}
and then the claim follows from \eqref{Wg-4}.

\begin{comment}

The calculation of the averages over the orthogonal group  is similar, slightly easier in fact, since no sign factors are involved. So we close this appendix with a unitary example, the proof of 
Lemma \ref{lem-Umoments} (iv). We have to integrate summands
of the form
$$ \overline{u}_{i_2 i_1} a_{i_2 i_3} u_{i_3 i_4}
b_{i_4 i_5} u_{i_6 i_5} c_{i_6 i_7} \overline{u}_{i_7 i_8}
d_{i_8 i_1}$$ over the unitary group. This is a constant factor times a unitary integral, and it is convenient to write the integrand
in the order $$ u_{i_3 i_4}\ u_{i_6 i_5}\
\overline{u}_{i_2 i_1}\ \overline{u}_{i_7 i_8}.$$ Now we have to consider all pairs in $\on{S}_2 \times \on{S}_2$. For instance,
the pair $(\on{id}, (12))$ leads to the conditions
$i_3 = i_2, i_7 = i_6, i_8 = i_4, i_5 = i_1$, hence to a prefactor
$$a_{i_2 i_2}\ b_{i_4 i_1}\ c_{i_6 i_6}\ d_{i_4 i_1},$$ and summing
over the these indices leads to a contribution
$$\tr(A) \tr(C) \tr(B D^t)\ \on{Wg}(\on{id}, (12)).$$

\end{comment}

%%%%%%%%%%%%%%%%%%%%%%%%%%%%%%%%%%%%%%%%%%%%%%
\subsection{The general Weingarten integration formulas}
\label{Wg-general}

This appendix collects the general results on the integration over the orthogonal, symplectic and unitary group, as they are given in (or can be deduced from) \cite{CS}, \cite{CSt} and \cite{Co} respectively. The following notation will be used. For $m, n \in \mathbb{N}$ denote by $\calf(m, n)$ the set of all maps from $\{ 1, \ldots, m\}$ to $\{ 1, \ldots, n\}$. If $\frm$ is a partition of $\{ 1, \ldots, m\}$, then $\calf(\frm, n)$ denotes the set of those elements of $\calf(m, n)$ which are constant on the blocks of $\frm$. Denote by $\Pp\Pp(k)$ the set of all pair partitions of $\{ 1, \ldots, k\}$. In particular, $\Pp\Pp(k) = \emptyset$ if $k$ is odd. If $k = 2l$ is even, we write $\Pp\Pp(k) \ni \frm = \{ \{m_{\nu}(\frm), n_{\nu}(\frm)\}\,|\, \nu = 1, \ldots, l\}$, where the numbering of the blocks is arbitrary, but $m_{\nu} < n_{\nu}$ holds for all $\nu$. For maps $\alpha\in \calf(k, \{0, 1\})$ from $\{ 1, \ldots, k\}$ to $\{0,1\}$, let us write $\alpha \in \cala(\frm, \{0, 1\})$ if for all $\nu$ holds  $\alpha(m_{\nu}(\frm)) = 1 - \alpha(n_{\nu}(\frm))$. 

\vspace{.2cm}

Now let us begin with the orthogonal case. If $N \ge l$, then the space of $\on{O}(N)$-invariants in $V^{\otimes 2l}$ admits a basis $\{\theta_{\frm}\,|\, \frm \in \Pp\Pp(2l)\}$ given by 
$$ 
\theta_{\frm} 
\;=\; 
\sum_{\phi \in \calf(\frm, N)} e_{\phi(1)} \otimes
\ldots \otimes e_{\phi(2l)}\;,
$$
where as above $e_i$, $i=1, \ldots, N$, is the standard orthonormal basis on $V=\CM^N$ which is furnished with the symmetric bilinear form $b(v,w)=v^tw$. The Weingarten function $\on{Wg}_{\on{O}(N)}(\frm, \frn)$ is the $(\frm, \frn)$-entry of the inverse of the Gram matrix of the extension of $b$ via  $b(v_1 \otimes \ldots \otimes v_{2l}, w_1 \otimes \ldots \otimes w_{2l})) = b(v_1, w_1) \cdot \ldots \cdot  b(v_{2l}, w_{2l})$ with respect to the basis  $\{\theta_{\frm}\,|\, \frm \in \Pp\Pp(2l)\}$. For example, on $\Pp\Pp(2)\times \Pp\Pp(2)$ one obtains  $\on{Wg}_{\on{O}(N)} = \frac{1}{N}$, while on $\Pp\Pp(4)\times \Pp\Pp(4)$ the orthogonal Weingarten function $\on{Wg}_{\on{O}(N)} $ is given by the matrix entries
$$
\frac{1}{N (N-1) (N+2)}\, 
\left( \begin{array}{lll}
N+1 & -1 & -1 \\ -1 & N+1 & -1 \\ -1 & -1 & N+1
\end{array} \right)
\;.
$$
The general orthogonal Weingarten integration formula now reads as follows.

%%%%%%%%%%%%%%%%%%%%%%%%%%%%%%%%%%%%%%%%%%%%%%
\begin{proposi}
\label{int_o} Let $\phi, \psi \in \calf(k, N)$. Then 
$$\int_{{\rm
O}(N)} \!\!dU\;
\prod_{j=1}^k \;U_{\phi(j), \psi(j)}
\;=\;
\sum_{\frm, \frn \in \Pp\Pp(k)} 1_{\calf(\frm, n)}(\phi)\
1_{\calf(\frn, n)}(\psi)\ \on{Wg}_{{\rm O}(N)}(\frm, \frn)
\;.
$$ 
%
%In particular, the integral vanishes if $k$ is odd.
\end{proposi}
%%%%%%%%%%%%%%%%%%%%%%%%%%%%%%%%%%%%%%%%%%%%%%

\vspace{.2cm}

Next let us consider the symplectic case, hence $V=\CM^{2N}$ with $a$ as in Section~\ref{App-sympl}. If $N \ge l$, then the space of $\on{SP}(2N)$-invariants in  $V^{\otimes 2l}$ admits a basis $\{\theta_{\frm}\,|\, \frm \in \Pp\Pp(2l)\}$, where 
$$ 
\theta_{\frm} 
\;=\; 
\sum_{\eta \in \calf(l, n)}
\sum_{\varepsilon \in \calf(l, \{ 0, 1\})}\ \otimes_{j=1}^{2l}
v(\eta, \epsilon, j)
\;,
$$
with
$$ 
v(\eta, \epsilon, j) 
\;=\; 
\left\{
\begin{array}{ll}
e_{\nu(\eta + n \epsilon)}\;, & \;\;{\rm if~} j = m_{\nu}\;,
\\
e_{\nu(\eta + n(1-\epsilon))}\ (-1)^{\nu \epsilon}\;, & \;\;
{\rm if~} j =
n_{\nu}.\end{array} \;.
\right.
$$
The Weingarten function $\on{Wg}_{\on{SP}(2N)}(\frm, \frn)$ is the $(\frm, \frn)$-entry of the inverse of the Gram matrix of the extension of $a$ (as above)  with respect to the basis  $\{\theta_{\frm}\,|\, \frm \in \Pp\Pp(2l)\}$. On $\Pp\Pp(2)\times \Pp\Pp(2)$ one obtains  $\on{Wg}_{\on{SP}(2N)} = \frac{1}{2N}$, and on $\Pp\Pp(4)\times \Pp\Pp(4)$ it is given by \eqref{Wg-4} above.

%%%%%%%%%%%%%%%%%%%%%%%%%%%%%%%%%%%%%%%%%%%%%%
\begin{proposi}
\label{int_sp} Let $\phi, \psi \in \calf(k, N),\ \alpha, \beta \in\calf(k, \{ 0, 1\}).$ Then
\begin{align*}
\int_{{\rm SP}(2N)} \!\!dU\;\prod_{j=1}^k\; U_{\phi(j) +
N\alpha(j), \psi(j) + N\beta(j)}
\, =
\sum_{\frm, \frn \in \Pp\Pp(k)} &  \on{Wg}_{{\rm
SP}(2N)}(\frm, \frn)\,
(-1)^{\sum_{\nu = 1}^l \alpha(m_{\nu}(\frm)) + \beta(m_{\nu}(\frn))}
\\
& 1_{\calf(\frm, 2N)}(\phi)\,
1_{\cala(\frm, 2N)}(\alpha)\,
1_{\calf(\frn, 2N)}(\psi)\,
1_{\cala(\frn, 2N)}(\beta)
\;.
\end{align*}
%In particular, the integral vanishes if $k$ is odd.
\end{proposi}
%%%%%%%%%%%%%%%%%%%%%%%%%%%%%%%%%%%%%%%%%%%%%%

Finally let us turn to the unitary case. Let $\on{S}_k$ denote the full symmetric  permutation group of the set $\{1, \ldots, k\}$, and $e_1^*, \ldots, e_N^*$ the dual to the basis $e_1, \ldots, e_N$ on $V=\CM^N$ w.r.t. the standard scalar product $c$.  If $N \ge k$, then the space of $\on{U}(N)$-invariants of $V^{\otimes k} \otimes (V^*)^{\otimes k}$ (with the contragredient representation on the dual space) admits a basis $\{ C_{\pi}\,|\, \pi \in \on{S}_k\}$ where 
$$
C_{\pi}
\; = \;
\sum_{\phi \in \calf(k, N)}
e_{\phi(\pi^{-1}(1))} \otimes \ldots \otimes 
e_{\phi(\pi^{-1}(k))} \otimes e^*_{\phi(1)} \otimes \ldots 
\otimes e^*_{\phi(k)}
\;.
$$
The Weingarten function $\on{Wg}_{\on{U}(N)}(\sigma, \tau)$ is the $(\sigma, \tau)$-entry of the inverse of the Gram matrix of the extension of the scalar product $c$ w.r.t.  the basis 
$\{ C_{\pi}\,|\, \pi \in \on{S}_k\}$. Then on $\on{S}_1 \times \on{S}_1$ one has $\on{Wg}_{\on{U}(N)} = \frac{1}{N}$, while on $\on{S}_2 \times
\on{S}_2$ the unitary Weingarten function is given by
$$ 
\frac{1}{N(N^2 - 1)} \left(\begin{array}{rr} 
N & -1 \\ -1 & N\end{array} \right)
\;.
$$

%%%%%%%%%%%%%%%%%%%%%%%%%%%%%%%%%%%%%%%%%%%%%%
\begin{proposi}
\label{int_u} Let $\phi, \psi, \phi\tra, \psi\tra \in \calf(k, N)$. Then
$$
 \int_{{\rm U}(N)} \!\!dU\;
 \prod_{j = 1}^k \;U_{\phi(j), \psi(j)}\;\overline{U}_{\phi\tra(j), \psi\tra(j)}
 \; =\;
\sum_{\sigma, \tau \in {\rm S}_k} \on{Wg}_{{\rm U}(N)}(\sigma, \tau)\ \delta(\phi, \sigma \phi\tra)\
\delta(\psi, \tau \psi\tra)
\;.
$$
\end{proposi}
%%%%%%%%%%%%%%%%%%%%%%%%%%%%%%%%%%%%%%%%%%%%%%

Let us conclude with a comment on higher order Weingarten functions. The definitions given here have the merit that they can be easily motivated in an invariant theoretic context. Their calculation, in particular, for higher orders, can be simplified by remarking, e.g., that $\on{Wg}_{\on{U}(N)}(\sigma,\tau)$ depends only on the cycle type of $\sigma \tau^{-1}$. This combined with harmonic analysis on the symmetric group leads to sophisticated and computationally feasible expansions of $\on{Wg}$, see \cite{CoMa} for the unitary and orthogonal cases. The recent preprint \cite{Mat} which contains an expansion in the symplectic case was posted only after the present paper had been submitted.

%%%%%%%%%%%%%%%%%%%%%%%%%%%%%%%%%%%%%%%%%%%%%%%%%%%%%%%%%%%%%%%%%%%%

\end{document}